\documentclass[12pt]{article}
\usepackage{epsfig,psfrag}
\usepackage{graphicx,amsfonts,amsmath,amssymb}
\usepackage[usenames]{color}
\def\Im{\mathrm{Im}}
\def\Re{\mathrm{Re}}

\begin{document}
\baselineskip 0.6cm
\newcommand{\gsim}{ \mathop{}_{\textstyle \sim}^{\textstyle >} }
\newcommand{\lsim}{ \mathop{}_{\textstyle \sim}^{\textstyle 3<} }
\newcommand{\vev}[1]{ \left\langle {#1} \right\rangle }
\newcommand{\bra}[1]{ \langle {#1} | }
\newcommand{\ket}[1]{ | {#1} \rangle }
\newcommand{\Dsl}{\mbox{\ooalign{\hfil/\hfil\crcr$D$}}}
\newcommand{\nequiv}{\mbox{\ooalign{\hfil/\hfil\crcr$\equiv$}}}
\newcommand{\nsupset}{\mbox{\ooalign{\hfil/\hfil\crcr$\supset$}}}
\newcommand{\nni}{\mbox{\ooalign{\hfil/\hfil\crcr$\ni$}}}
\newcommand{\EV}{ {\rm eV} }
\newcommand{\KEV}{ {\rm keV} }
\newcommand{\MEV}{ {\rm MeV} }
\newcommand{\GEV}{ {\rm GeV} }
\newcommand{\TEV}{ {\rm TeV} }

\def\diag{\mathop{\rm diag}\nolimits}
\def\tr{\mathop{\rm tr}}

\def\Spin{\mathop{\rm Spin}}
\def\SO{\mathop{\rm SO}}
\def\O{\mathop{\rm O}}
\def\SU{\mathop{\rm SU}}
\def\U{\mathop{\rm U}}
\def\Sp{\mathop{\rm Sp}}
\def\SL{\mathop{\rm SL}}

\def\change#1#2{{\color{blue}#1}{\color{red} #2}\color{black}\hbox{}}


\begin{titlepage}

\begin{flushright}
\end{flushright}

\vskip 2cm
\begin{center}
{\Large \bf D--terms on the resolved conifold}
\vskip 1.2cm
{\bf Keshav Dasgupta, Paul Franche, Anke Knauf and James Sully}

\vskip 0.4cm

{\it Rutherford Physics Building, McGill University, Montreal, QC H3A 2T8, Canada}

\vskip 0.2cm

{\tt keshav, franchep, knauf, sullyj@hep.physics.mcgill.ca}

\vskip 1.5cm

\abstract{We derive a novel deformation of the warped resolved conifold background with supersymmetry breaking ISD (1,2) fluxes by adding D7--branes to this type IIB theory. We find spontaneous supersymmetry breaking without generating a bulk cosmological constant. In the compactified form, our background will no longer be a Calabi--Yau manifold as it allows a non--vanishing first Chern class. In the presence of D7--branes the (1,2) fluxes can give rise to non-trivial D-terms. We study the Ouyang embedding of D7--branes in detail and find that in this case the D--terms are indeed non-zero. In the limit when we approach the singular conifold, the D--terms vanish for Ouyang's embedding, although supersymmetry appears to be broken.

We also construct the F-theory lift of our background and demonstrate how these IIB (1,2) fluxes lift to non--primitive (2,2) flux on the fourfold. The seven branes correspond to normalisable harmonic forms. 
We briefly sketch a possible way to attain an inflaton potential in this background 
once extra D3--branes are introduced and  
point out some possibilities of 
restoring supersymmetry in our background that could in principle be used as the end point of the
inflationary set-up. In
a companion paper we will analyse in details the inflationary dynamics in this background.} 

\end{center}
\end{titlepage}

\tableofcontents
\pagebreak

\section{Introduction}
\def\theequation{\arabic{section}.\arabic{equation}}
\setcounter{equation}{0}

\subsection{Motivation}

Our motivation in studying the warped resolved conifold with soft supersymmetry breaking is to come a step closer to a consistent string theory background that can be used to study inflation. 
Current D--brane inflation models (e.g. \cite{KKLMMT,bdkm, axel, BCDF}) are usually embedded in a particular type IIB string theory setup that has become known as the ``warped throat''. It is a background on which fluxes create a strongly warped Calabi--Yau geometry via their backreaction on the metric. The Calabi--Yau in question is taken to be the conifold or its cousin the deformed conifold, in which the tip of the throat is non--singular. Placing an anti--D-brane at the bottom of the throat and a D-brane at some distance from it, breaks supersymmetry. Consequently, the D-brane is attracted towards the bottom of the throat with the inter--brane distance serving as the inflaton. As has been pointed out in a variety of papers \cite{KKLMMT, bdkm}, it is very hard to achieve slow roll in these models. 

As an alternative one can break supersymmetry spontaneously by turning on appropriate fluxes, e.g. instead of lifting the potential with an anti--D-brane, one can turn on D--terms. (This idea was put forward in \cite{quevedo}, but needed some corrections \cite{achucarro, fernando}. In short, one can only generate D-terms in a non-susy theory, i.e. if there are also F-terms present \cite{nilles}.)

There has been much interest in D--terms coming from string theory \cite{uranga, lust, hans, haacklust, berglund}
both for particle phenomenology and cosmological applications. D--terms can generically be created by non--primitive flux on D--brane worldvolumes. It turns out, however, that in the case of only D3--branes, the D--terms will vanish in the vacuum \cite{uranga}. Even with D7--branes and D3/D7 setups, the cycles wrapped by the branes need to fulfill non--trivial topological conditions to achieve a D-term uplifting \cite{hans}. Although D-brane inflation mostly considers D3--branes, D7--branes have been established as a key ingredient for moduli stabilisation. Non--perturbative effects (gaugino condensation) on their worldvolume allow the stabilization of the overall radial modulus.

In light of this knowledge, we propose a background that breaks supersymmetry, but still solves the supergravity equations of motion. It contains D7--branes, which allow for the creation of D--terms. With cosmological applications in mind, this background is a ``relative'' of the warped throat, i.e. it looks asymptotically like a conifold, but has a different behaviour near the tip. The key ingredient is the blow--up of a 2--cycle (in contrast to the 3--cycle of the deformed conifold), which will introduce non--primitive flux into the theory. This flux still solves the equation of motion as it is imaginary self--dual (ISD). Generically, such a flux cannot exist on a compact Calabi--Yau. We therefore have to generalise our manifold to some non--CY compactification, or keep the whole setup non--compact. For simplicity, we will follow the latter approach, giving some speculations about what a consistent non--CY compactification might induce.

\subsection{The background}

The simplest ``throat'' studied so far is the singular conifold, a warped flux background known as the Klebanov--Tseytlin (KT) solution \cite{kt}. The singularity at the tip of the conifold can be smoothed out in two different ways: by blowing up a 3--sphere (the deformed conifold) or by blowing up a 2--sphere (the resolved conifold). Both these manifolds are still Calabi--Yau. These particular backgrounds, with added fluxes, have been studied by Klebanov--Strassler (KS) \cite{ks} and Pando Zayas--Tseytlin (PT) \cite{pt} respectively. 

On the other hand, one could imagine a more general background that allows for both blown--up 2-- and 3--cycles. The ``resolved warped deformed conifold'' can be interpreted as such a manifold. It was introduced \cite{DKS} as an interpolating solution between the KS and Maldacena--Nunez (MN) solutions (see also \cite{papatseyt, butti}). It is not a CY anymore, but an SU(3) structure manifold. Apart from the blown--up 2-cycle, there is another interesting feature: the background exhibits a running dilaton, in contrast to the KT, KS or PT solutions on warped CY's with constant dilaton. Placing a D3--brane in this background will result in a force due to this running dilaton. This does not mean that the resolved warped deformed conifold breaks supersymmetry, but rather that the D3 oriented along Minkowski space does not preserve the same subset of supercharges.
There is another source of a running dilaton that will be of interest to us: D7--branes. Their behaviour will be determined by the particular embedding we choose for the D7. 

The most general ``throat'' background, taken to be the resolved warped deformed conifold, has the metric 
\begin{eqnarray}\label{resdefmetric}
  ds^2   &=& F_3~ dr^2 + F_4(d\psi + \cos\theta_1\, d\phi_1 + \cos\theta_2\, d\phi_2)^2\\ 
  & + &  F_1\, \big(d\theta_1^2+\sin^2\theta_1\,d\phi_1^2\big) 
     +~ F_2\, \big(d\theta_2^2+\sin^2\theta_2\,d\phi_2^2\big) \nonumber\\ 
  & + &  2b \bigg[\cos\psi \big(d\theta_1 d\theta_2 + \sin\theta_1 \sin\theta_2 d\phi_1 d\phi_2\big)  
- \sin\psi \big(\sin\theta_2 d\phi_2 d\theta_1 - \sin\theta_1 d\phi_1 d\theta_2\big)\bigg]\nonumber
\end{eqnarray}
where the coefficients $F_i, b$ are functions of the radial coordinate $r$, $(\theta_i,\phi_i)$ parameterise two 2--spheres, and $\psi=0\ldots 4\pi$ is a U(1) fibration over those spheres. The commonly known backgrounds are found in the limits:
\begin{itemize}
 \item singular conifold: $F_1=F_2$ and $b=0$, i.e. both 2-spheres have equal radii (and shrink to zero size as $r\to 0$), the cross--terms in the third line in \eqref{resdefmetric} are absent
 \item deformed conifold: $F_1=F_2$ and $b\ne 0$, i.e. both 2-spheres have equal radii, but the U(1) shift symmetry is broken due to the more complicated fibration in the third line
 \item resolved conifold: $F_1\ne F_2$ and $b=0$, i.e. the 2--spheres have unequal size (this corresponds to the breaking of a discrete $\mathbb{Z}_2$ exchanging both) and the third line in \eqref{resdefmetric} is absent
\end{itemize}
For a complete definition of the functions $F_i$ we refer the reader to \cite{DKS, candelas, ankenon}. They are of course more restricted than outlined above in order to guarantee an SU(3) holonomy or SU(3) structure.
In \cite{BCDF}, the limit  
\begin{equation}\label{oldpap}
  F_1 ~ \approx ~ F_2 ~ = ~ \frac{r^2}{6}, ~~~~~ b ~ \to ~ 0, ~~~~~~ F_3 ~ = ~ 1, ~~~~~ F_4 ~=~ \frac{r^2}{9}
\end{equation}
was employed. 
In this limit the background becomes a (non-compact) singular conifold, and one can add D7 branes using the technique 
discussed in \cite{ouyang}. This is the simplest choice and works well in the situation when we are far from the 
tip of the throat and the resolution parameter (the size of the 2--sphere that remains finite) is very small.
Here, we intend to go beyond this simplification. However, the resolved warped deformed conifold is difficult to study, mostly because it is not a CY. We therefore choose the simplest approximation that captures the essential feature of the blown--up 2--cycle: We choose to restrict ourselves to the resolved conifold.

We will turn on fluxes (or rather borrow them from the PT solution \cite{pt}) that break supersymmetry because they are not only of cohomology type (2,1), but also (1,2). This is not possible on a compact CY. (1,2) flux can only be ISD if it is of the form $J^{1,1}\wedge \bar{m}^{0,1}$, for some antiholomorphic 1--form $\bar m$ ($J$ is the K\"ahler form). This would require a nontrivial one--cycle, so the first Chern class cannot be zero anymore. This argument breaks down for non--compact manifolds, as Poincar\'e duality fails. For the compact cycles there is still a correspondence between homology and cohomology though. In a consistent compactification, one therefore has to change the background as to not be conformally CY, or to glue it onto a compact bulk in such that the entire compactification manifold is no longer CY. This would lead us beyond the case of conformal CY with flux compactifications examined in \cite{keshavsav} or GKP \cite{GKP}, and is beyond the scope of this work. In section \ref{ptbg} we will review the PT background and explain why it already breaks supersymmetry. It will be shown, however, that this does not lead to uplifting as the cosmological constant remains zero (this is explained in section \ref{cc}). Only after we embed D7--branes in this background (see section \ref{ouyangbed}) we can observe the D--terms and uplift our potential. This calculation is performed in section \ref{dterms}.

An alternative view on the problem is given by lifting the whole scenario to F--theory in section \ref{ftheory}. We resolve some of the subtleties associated with the lift, namely the existence of seven branes, the existence of non-primitive 
fluxes and the existence of a compact geometry. We show that the type IIB seven branes are directly related to 
certain normalisable harmonic forms and we construct them explicitly. These forms are the ones that 
contribute to the second cohomology of the compact manifold. We argue that the compact geometry cannot be a Calabi-Yau
manifold by demonstrating that the first Chern class does not vanish. We show that the non-K\"ahlerity can be attributed to 
the existence of a three form in the dual type IIA theory. We also argue that the IIB (1,2) forms can combine with the 
non-K\"ahlerity to form a unique (2,2) form in the M-theory lift of our background. In section \ref{cosmo} we sketch a possible 
inflationary model from our scenario, and point out a process of restoring supersymmetry at the end of inflation. 
In a companion paper we will analyse detailed inflationary dynamics in this background. 

\section{The IIB picture: D7--branes on the resolved conifold}\label{iib}
\setcounter{equation}{0}

In the following we describe the basic geometry of the resolved conifold background and then show how branes and susy--breaking fluxes can be 
consistently added without violating the equations of motion.  

\subsection{The warped resolved conifold with fluxes}\label{ptbg}

Similar to the Klebanov--Strassler model, a warped geometry can be created by fluxes in the \emph{resolved} conifold background, see appendix \ref{rescone} for a discussion of this geometry and definition of coordinates. 
The full supergravity solution for the resolved conifold was derived by Pando--Zayas and Tseytlin \cite{pt} (PT) and includes non--trivial RR and NS flux with constant dilaton. It can be understood as placing a stack of fractional D3--branes (i.e. D5--branes that wrap a 2--cycle) in this background. The ten--dimensional metric is found to be
\begin{equation}
\label{pzmet}
 ds_{10}^2=h^{-1/2}(\rho)\,\eta_{\mu\nu}dx^\mu dx^\nu +h^{1/2}(\rho)\,ds_6^2\,,
\end{equation}
where $ds_6^2$ refers to the resolved conifold metric given by
\begin{eqnarray}\nonumber
  ds^2_{6} & = & \kappa(\rho)^{-1}\,d\rho^2 + \frac{\kappa(\rho)}{9}\,\rho^2\big(d\psi+\cos\theta_1\,d\phi_1
    +\cos\theta_2\,d\phi_2\big)^2 \\
  & &+ \frac{\rho^2}{6}\,\big(d\theta_1^2+\sin^2\theta_1\,d\phi_1^2\big) 
    +\frac{\rho^2+6a^2}{6}\,\big(d\theta_2^2+\sin^2\theta_2\,d\phi_2^2\big)\,.
\end{eqnarray}
Note that as $\rho\to 0$, the $(\theta_2,\phi_2)$ sphere remains finite, whereas for the singular conifold both $(\theta_i,\phi_i)$ spheres scale with $\rho^2/6$. The parameter $a$ is called the resolution parameter because it determines the size of the resolved 2--sphere. This asymmetry in the geometry also determines an asymmetry in the flux on the 2--cycles and is the source of supersymmetry breaking.
The 3--form fluxes in this background are\footnote{There is a typo in eq. (4.3) in \cite{pt}, concerning the sign of $F_3$.}
\begin{eqnarray}\label{fluxres}
  H_3 & = & d\rho\wedge[f'_1(\rho) \,d\theta_1 \wedge \sin\theta_1\,d\phi_1+f'_2(\rho)\,d\theta_2\wedge\sin\theta_2\,d\phi_2]  \\
  F_3 & = & P e_{\psi}\wedge (d\theta_1\wedge \sin\theta_1\,d\phi_1- d\theta_2\wedge\sin\theta_2\,d\phi_2 )
\end{eqnarray}
and the self--dual 5--form flux is given by
\begin{equation}
F_5 = {\cal F}+* {\cal F}\ , \quad  \ \ \ \ {\cal F} = K(\rho)\,e_{\psi}\wedge d\theta_1 \wedge\sin\theta_1\,d\phi_1 \wedge d\theta_2\wedge \sin\theta_2\,d\phi_2 \ ,
\end{equation}
where
\begin{eqnarray}\label{deffs}\nonumber
  f_1(\rho) & =& \frac{3}{2} g_s P \ln (\rho^2+9a^2)  \\
  f_2(\rho) & =&   \frac{1}{6} g_s P \bigg(  \frac{36a^2}{\rho^2} - \ln [\rho^{16} (\rho^2+9a^2)] \bigg)  \\ \nonumber
  K(\rho)   & =&  Q - \frac{1}{3}  g_s P^2  \bigg(  \frac{18a^2}{\rho^2}- \ln [\rho^{8} (\rho^2+9a^2)^5] \bigg) 
\end{eqnarray}
and where P is proportional to the number of fractional D3-branes and Q proportional to the number of regular D3-branes, and both are proportional to $\alpha^\prime$.

It was pointed out in \cite{cvetgiblupo} and confirmed in \cite{ankenon} that this solution breaks supersymmetry. The reason lies in the fact that the 3--form flux has not only a (2,1), but also a (1,2) part. It is, nevertheless, a supergravity solution because the 3--form flux $G_3=F_3-iH_3$ obeys the imaginary self--duality condition $*_6 G_3=iG_3$. Supersymmetry further requires $G_3$ to be of type (2,1) and primitive \cite{granapol, kst},
i.e. that it satisfy $G_3\wedge J=0$.

Let us briefly review the argument. Using \eqref{resvielb} we can rewrite the 3--form flux in terms of vielbeins
\begin{equation}
G_3 \,=\, -\frac{18 P}{\rho^3\sqrt{\kappa}}\,(e_2\wedge e_3\wedge e_4+i\, e_1\wedge e_5\wedge e_6)
 +\frac{18P\,(e_2\wedge e_5\wedge e_6+i\, e_1\wedge e_3\wedge e_4)}{\rho\sqrt{\rho^2+6a^2}\sqrt{\rho^2+9a^2}}\,.
\end{equation}
The vielbein notation is extremely convenient to see that this flux is indeed imaginary self-dual. The Hodge dual is simply found by
\begin{equation}\nonumber
*_6 (e_{i_1}\wedge e_{i_2}\wedge\ldots\wedge e_{i_k}) = \epsilon_{i_1 i_2\ldots i_k}^{\phantom{i_1 i_2\ldots i_k}i_{k+1}\ldots i_6}\,e_{i_{k+1}}\wedge \ldots \wedge e_{i_6}
\end{equation}
and does not involve any factors of $\sqrt{g}$. We use the convention that $\epsilon_{123456}=\epsilon_{123}^{\phantom{123}456}=1$.
With the complex structure \eqref{cs} the PT flux becomes
\begin{eqnarray}\nonumber
  G_3 &=& \frac{-9 P}{\rho^3\sqrt{\rho^2+6a^2}\sqrt{\rho^2+9a^2}}\,\Big[(\rho^2+3a^2)\,(E_1\wedge E_2\wedge \overline E_2 
    - E_1\wedge E_3\wedge\overline E_3)\\
  & & \quad\phantom{\rho^3\sqrt{\rho^2+6a^2}\sqrt{\rho^2+9a^2}}+ 3a^2\,(E_2\wedge\overline E_1\wedge\overline E_2 + E_3\wedge\overline E_1\wedge\overline E_3)\Big]\,.
\end{eqnarray}
We make several observations: This flux is neither primitive\footnote{Since $J=\frac{\imath}{2}\,\sum_i(E_i\wedge\overline E_i)$ it
follows immediately that $J\wedge G_3$ has a nonvanishing $E_2\wedge E_3\wedge \overline{E_1}\wedge \overline{E_2}\wedge\overline{E_3}$ part that is proportional to $a^2$.} nor is it of type (2,1). It has a (1,2) \em
and \em a (2,1) part, which cannot be avoided by a different choice of complex structure. Consequently, this flux indeed breaks supersymmetry. 

We also observe that, in the limit $a\to 0$, the (1,2) part vanishes, the flux becomes primitive, and we recover
the singular conifold solution. This indicates that the resolution forbids a supersymmetric supergravity solution, i.e. the blow--up of a nontrivial 2--cycle in a conifold geometry can lead to supersymmetry breaking. We will exploit this fact to our advantage.

\subsection{The scalar potential and supersymmetry}\label{cc}

We have just argued that the non-primitive (1,2) flux breaks supersymmetry. One might therefore wonder if it can be used to uplift our potential to a positive vacuum. The answer is no because the scalar potential always remains zero when the flux is ISD, regardless of whether or not the vacuum breaks supersymmetry. Let us explain this in more detail (see also appendix (A.2) of \cite{GKP} and \cite{kst}). First we would like to remind the reader that the ISD requirement for $G_3$ stems from the SuGra equations of motion in compactifications on conformal CY's, as first pointed out by \cite{keshavsav, BeckerM} and later on elaborated by GKP \cite{GKP}, whereas the explicit susy variations lead to $J\wedge G_3=0$ (primitivity) and $G_3$ being purely (2,1). So the PT flux breaks susy ``in two ways'', by being (1,2) and by being non--primitive, which is actually one and the same statement for ISD fluxes.

The scalar potential of $\mathcal{N}=1$ 4d supergravity 
can be derived by direct dimensional reduction of the IIB SuGra action. It is induced by the flux kinetic term
\begin{equation}
  S_G \,=\, -\frac{1}{4\kappa_{10}^2}\int \frac{G_3\wedge *\overline{G}_3}{\Im\,\tau}\,,
\end{equation}
where the Hodge star is taken on the internal manifold, so this integral runs over the six internal dimensions.
This can be rewritten as a potential plus a topological term, if we split $G_3$ in its ISD and anti-ISD part
\begin{eqnarray}\nonumber
  && G_3 = G^{\rm ISD}+G^{\rm AISD}\,,\qquad\qquad G^{\rm (A)ISD} 
 \,\equiv \, \frac{1}{2}\big(G_3\pm i *G_3\big)\\
  && *G^{\rm ISD} =  iG^{\rm ISD}\,, \qquad\qquad\qquad  *G^{\rm AISD} \,=\,  -iG^{\rm AISD}\,.
\end{eqnarray}
Then this part of the action becomes
\begin{eqnarray}\nonumber
  S_G &=& -\frac{1}{2\kappa_{10}^2}\int \frac{G^{\rm AISD}\wedge *\overline{G}^{\rm AISD}}{\Im\tau}+\frac{i}{4\kappa_{10}^2}\int 
    \frac{G_3\wedge \overline{G}_3}{\Im\tau}\\[1ex]
  &=& -V -N_{\rm flux}\,.
\end{eqnarray}
The second term is topological and independent of the moduli. In a compact setup it will be cancelled by the localised charges,
if we use the tadpole cancellation condition $\int H_3\wedge F_3=-2\kappa_{10}^2T_3 Q_3^{\rm loc}$. (The D7--branes also carry an effective D3--charge given by $-\chi(X)/24$, the Euler character of the corresponding F--theory 4--fold.) This condition is of course relaxed in a non--compact space, but we want to keep the point of view that we can consistently compactify our background in an F--theory framework. The potential for the moduli is given by the anti-ISD fluxes only\footnote{For a more precise treatment that also includes warping, the Einstein term and the $F_5$ flux term see \cite{dewolfe}. The qualitative result remains unchanged. It was actually shown that the GVW superpotential is not influenced by warping.}
\begin{equation}\label{aisdpot}
  V \,=\, \frac{1}{2\kappa_{10}^2}\int \frac{G^{\rm AISD}\wedge *\overline{G}^{\rm AISD}}{\Im\tau}\,.
\end{equation}
This means that the potential vanishes identically for ISD flux and the ensuing condition $*G_3=iG_3$ fixes almost all moduli, namely complex structure moduli and dilaton.

If the basis of the complex structure moduli space is given by the holomorphic 3-form $\Omega$ (which is AISD) and $h^{2,1}$ primitive ISD (2,1) forms $\chi_i$, the flux $G_3$ is expanded in this basis. Upon this expansion, the scalar potential takes a form that only depends on the coefficients of the expansion of the anti--ISD part
\begin{equation}\label{expandG}
  G_3^{\rm AISD} \,=\, g_1\, \Omega + g_2^i\,\bar \chi_i
\end{equation}
and becomes
\begin{equation}\label{vexpand}
  V \,=\, \frac{i\int G_3\wedge\overline{\Omega}\int \overline{G}_3\wedge\Omega+\int G_3\wedge \chi_i\int\overline{G}_3\wedge \overline{\chi}^i}{2\,\Im\tau\,\kappa_{10}^2\int \Omega\wedge\overline{\Omega}}\,.
\end{equation}
This is identical to the standard scalar potential of $\mathcal{N}=1$ 4d supergravity in terms on the superpotential $W$ and the K\"ahler potential $\mathcal{K}$
\begin{equation}\label{sugrapot}
  V \,=\, e^{\mathcal{K}}\left(\sum_\alpha\vert D_\alpha W\vert^2-3\vert W\vert^2\right)\,,
\end{equation}
if the superpotential is the usual Gukov--Vafa--Witten \cite{GVW} potential
\begin{equation}\label{gvwsuppot}
  W \,=\, \int G_3\wedge \Omega
\end{equation}
and the K\"ahler potential is given by $\mathcal{K}=-\log(-i\int \Omega\wedge\bar\Omega)-\log[-i(\tau-\bar\tau)]-3\log[-i(\sigma-\bar\sigma)]$, where $\sigma$ is the K\"ahler modulus associated with the overall volume of the Calabi--Yau. 
The (2,1) forms $\chi_i$ enter through the derivative of $\Omega$, because the derivative of $\Omega$ with respect to a complex structure parameter $z_j$ has a (3,0) and a (2,1) part (see e.g. \cite{ossa})
\begin{equation}\label{derivomega}
  \frac{\partial \Omega}{\partial z_j} \,=\, k_j(z,\bar z)\Omega^{(3,0)}+\chi_j^{(2,1)}\,.
\end{equation}
In \eqref{sugrapot} the index $\alpha$ runs over all K\"ahler moduli $k_a$, complex structure moduli $z_i$ and the dilaton $\Phi$. The K\"ahler covariant derivate is $D_\alpha W=\partial_\alpha W+W\,\partial_\alpha \mathcal{K}$. For no--scale models one finds a cancellation between the covariant derivatives w.r.t. the K\"ahler moduli against the last term, so that
\begin{equation}
   V \,=\, e^{\mathcal{K}}\sum_i\vert D_i W\vert^2\,,
\end{equation}
where now $i$ only runs over the complex structure moduli and $\Phi$ only. It is therefore easy to see that even a minimum with $V=0$ can have broken supersymmetry, as $D_{k_a} W$ can be nonvanishing.

Now let us turn to the question why the non--susy (1,2) flux does not lead to uplifting. It is ISD, so obviously the potential \eqref{aisdpot} remains zero. But how can we understand this from the point of view of the SuGra potential as expressed in \eqref{sugrapot}? 
Clearly, there is no F--term associated to derivatives w.r.t. the K\"ahler parameter or the dilaton, as the superpotential \eqref{gvwsuppot} does not depend on them. But what about an F--term $D_{z_j}W$?
Let us for a moment assume we are still talking about a CY, although (1,2) ISD flux cannot exist on a compact CY. So we still assume our moduli space to be parameterised by $\Omega$ and $\chi_i$. Let us furthermore assume the superpotential is still given by \eqref{gvwsuppot}. Then it is easy to see that there could be a non--vanishing derivative of $W$ w.r.t. a complex structure parameter. Using \eqref{derivomega} one finds
\begin{equation}
  \partial_{z_i} W \,=\, k_i(z,\bar z)\,W+\int G_3\wedge \chi_i^{(2,1)}\,,
\end{equation}
which could be nonvanishing for $G_3$ of type (1,2). But (1,2) flux can only be ISD if it is proportional to the K\"ahler form, $G^{(1,2)}=J^{(1,1)}\wedge \bar m^{(0,1)}$, so this becomes
\begin{equation}
  \partial_{z_i} W \,=\,\int J^{(1,1)}\wedge \bar m^{(0,1)}\wedge \chi_i^{(2,1)} \,=\, 0
\end{equation}
when we use the fact that $\chi_i$ is primitive, $J^{(1,1)}\wedge\chi_i^{(2,1)} \,=\, 0$. If there is no (0,3) part present, $W$ vanishes identically and 
\begin{equation}
  D_{z_i} W \,=\,  \partial_{z_i} W+W\,\partial_{z_i}\mathcal{K} \,=\,0\,,
\end{equation}
so all F--terms vanish in our setup. Note that in the non--compact scenario the term $-3\vert W\vert^2$ is absent (we neglected $M_P$ in above formulae). However, our argument does not depend on the no--scale structure of the model. $W$ is identically zero, because we don't have any (0,3) flux turned on, and all F-terms vanish individually.

This discussion has two weak points: First of all, we can no longer assume our moduli space is only parameterised by $\Omega$ and $\chi_i$ if we allow for a (1,2) flux. Once we compactify, there has to be a basis for the one--form $m^{(1,0)}$ as well (for simplicity of the argument let us assume there is only one such 1--form in the following). This would modify the derivative of $\Omega$, the natural guess respecting the  (3,0)+(2,1) structure\footnote{In the case of a complex manifold, the original derivation \cite{ossa} holds and \eqref{newdomega} would not acquire an extra term.} being 
\begin{equation}\label{newdomega}
  \frac{\partial \Omega}{\partial_{z_j}} \,=\, k_j(z,\bar z)\Omega^{(3,0)}+\chi_j^{(2,1)}+\nu_j\, J^{(1,1)}\wedge m^{(1,0)}\,.
\end{equation}
If we keep using the GVW superpotential, we get an additional term
\begin{equation}
  \partial_{z_j} W \,=\,  \int G_3\wedge (\nu_j\, J^{(1,1)}\wedge m^{(1,0)}) \,=\, \int J^{(1,1)}\wedge \bar 
    m^{(0,1)}\wedge \nu_j\, J^{(1,1)}\wedge m^{(1,0)}\,,
\end{equation}
which will in general be non--zero for the type of $G_3$ flux we have turned on. However, the superpotential will also change since we have to expand $G_3$ in this new basis as well. Equation \eqref{expandG} changes to
\begin{equation}
  G_3^{\rm AISD} \,=\, g_1\, \Omega + g_2^i\,\bar \chi_i + g_3\, J\wedge \bar m\,.
\end{equation}
Plugging this into the scalar potential \eqref{aisdpot} does not give \eqref{vexpand}, but additional terms due to $\bar m$. To bring this into the form of the standard SuGra F--term potential we would need to know the metric on the new moduli space, which does not correspond to a CY anymore. Finding the relevant moduli space would allow one to see how $W$ changes. It is likely that it will contain terms with $J$, and thus will introduce a dependence on K\"ahler structure moduli. This breaks the no--scale structure and we have to re--examine the cancellation between $D_{k_a} W$ and $W$. Regardless, we know that the combination $\sum_\alpha |D_\alpha W|^2-3|W|^2$ has to vanish, as \eqref{aisdpot} remains valid. ISD flux cannot give a non--zero potential.

In addition, it is worth noting that we may have to modify the superpotential as to include a term enforcing primitivity. In the compact CY setting this is already taken care of, because an ISD (2,1) form is always primitive. The ISD (1,2) form, on the other hand, is not. If we allow for this type of flux, we should introduce a term that reproduces the primitivity condition as a susy condition $DW=0$. This was already considered in an M/F--theory context \cite{GVW}, where it was conjectured that
\begin{equation}\label{tildew}
  \widetilde{W} \,=\, \int J\wedge J\wedge G_4\,.
\end{equation}
Then $D_J\widetilde{W}=0$ leads to the primitivity condition $J\wedge G_4=0$ for the 4-form flux on the 8--manifold. It is not obvious how this term reduces to type IIB. It will not give rise to a superpotential, but rather to a D--term, as it depends on the K\"ahler moduli and not the complex structure moduli.
For a  $K3\times K3$ orientifold, the dimensional reduction of $\widetilde{W}$ has been carried out \cite{lust} and the result agrees with that obtained in type IIB from a D7--worldvolume analysis \cite{hans}. Also in the F--theory setup, only the non--primitive fluxes on the D7--branes create a D--term in the effective four--dimensional theory. We can therefore safely conclude that the supersymmetry breaking due to the (1,2) flux will not be visible in the scalar potential that appears from the reduction of the IIB bulk action. 

There is also an enlightening discussion in \cite{haack} where it was illustrated that, from an F--theory point of view, a flux of type (0,4), (4,0) or proportional to $J\wedge J$ can break supersymmetry without generating a cosmological constant. It is the latter case that corresponds to non--primitive ISD flux in IIB. We do not have an explicit map between these two types of fluxes, but we give some arguments in section \ref{lift}. It should be clear that ISD flux lifts to self--dual flux in F-theory and that the non-primitivity property is preserved in this lift.

To summarise, the supersymmetry breaking associated to non--primitive (1,2) fluxes will not give rise to an F--term uplift, as the scalar potential generated by the flux in the IIB bulk action remains zero, so does the superpotential if we rely on the CY property of the resolved conifold. We can, however, in the spirit of KKLMMT allow a non--vanishing $W_0$ that is created by fluxes in the compact bulk that is glued to the throat. It does not appear in the scalar potential because of the no--scale structure of these models (but it will, once the no--scale structure is broken by non--perturbative effects or because the superpotential is not simply the one from GVW \cite{GVW} anymore). The (1,2) flux gives rise to an ``auxiliary D--term'' \cite{kst}, which is absent in the 4d scalar potential but can be understood as an FI--term from an anomalous $U(1)$ on the D7 worldvolume (the pullback of the B-field on the D7 worldvolume enters into the DBI action). Let us therefore turn to the question how to embed a D7 in the resolved conifold background; we will then turn to the computation of the D--terms in section \ref{dterms}.

\subsection{Ouyang embedding of D7--branes on the resolved conifold}\label{ouyangbed}

We consider now that addition of D7--branes to the PT background. 
In \cite{ouyang}, a holomorphic embedding of D7--branes into the singular conifold background was presented. Such an embedding is necessary to preserve supersymmetry on the submanifold, although not alone sufficient (complete BPS conditions are found in \cite{gomis, Marino:1999af}). The particular holomorphic embedding chosen in \cite{ouyang} is described by
\begin{equation}
  z \,=\, \mu^2\,,
\end{equation}
where $z$ is one of the holomorphic coordinates defined in \eqref{holocoord}. Although we already know that the PT background breaks supersymmetry, we will use precisely the same embedding (we consider only $\mu=0$ for simplicity). It is worth emphasising that this embedding, first considered on the singular conifold, remains holomorphic on the resolved conifold (details are found in Appendix \ref{embed}). As a consistency check we should always be able to recover the original singular solution in the limit $a\to 0$. This singular solution from \cite{ouyang} is actually not supersymmetric, though one might have expected otherwise. The embedding is holomorphic, but supersymmetry requires in addition that the pullback of the flux is (1,1) and primitive on the cycle wrapped by the D7. The latter condition is not met by the singular Ouyang embedding in \cite{ouyang}. It might be possible to restore supersymmetry by turning on appropriate gauge flux\footnote{P. Ouyang, G. Shiu et al, work in progress.}. However, as we will demonstrate in section \ref{dterms}, this susy breaking in \cite{ouyang} does not manifest itself in a D--term.

The D7--brane induces a non--trivial axion--dilaton
\begin{equation}\label{dilbehavior}
  \tau \,=\,  \frac{i}{g_s}+\frac{N}{2\pi i} \log z\,,
\end{equation} 
where $N$ is the number of embedded D7-branes. As pointed out in \cite{BCDF}, there is 
an additional running of the dilaton when the two--cycle in the ``resolved warped deformed conifold'' is blown up.
However, as we focus on the limit where the geometry looks like the resolved conifold (i.e. $b \to 0$ in \eqref{resdefmetric}), we recover the PT supergravity solution, which has a constant dilaton. We will therefore concentrate on the running of the dilaton \eqref{dilbehavior} as generated by the D7--brane embedding. This running dilaton was not taken into account by \cite{bdkm}, where the D7 is embedded in the singular conifold and a D3--brane is attracted towards an anti--D3 at the bottom of the throat. The given reasoning is that the dilaton contribution should be exactly cancelled by a change in geometry when approaching the supersymmetric limit (if the D7--brane embedding is supersymmetric and the D3--brane preserves the same supersymmetry, the scenario has to be stable when the susy--breaking anti--D3 is removed). Our setup, on the other hand, is non--supersymmetric from the start and therefore we are not led to conclude that the running of the dilaton should vanish from a similar line of argument. It will, however, be suppressed by the susy breaking scale. For a viable inflationary scenario one should rather use the resolved warped deformed conifold; its running dilaton will be the primary reason for a D3 to move towards the tip\footnote{Such a scenario has been studied in \cite{BCDF}, where the running dilaton due to a blown--up 2--cycle was parameterized by $\delta N(a)\,\log z$, where $a$ is a small resolution. This analysis was based on the original Ouyang embedding \cite{ouyang}, which we will now reconsider for the resolved conifold.}. In this section we simply want to study the backreaction of the dilaton onto the background.

We determine the change the dilaton induces in the other fluxes and the warp factor at linear order $g_sN$, see appendix \ref{embed} for details of the calculation. We neglect any backreaction on the geometry beyond a change in the warp factor, i.e. we will assume the manifold remains a conformal resolved conifold. A distortion of the conifold with Ouyang embedding has been studied in e.g. \cite{nunez}, where the D7--branes are smeared over the angular directions, such that the dilaton does not exhibit the behaviour \eqref{dilbehavior}, but runs as $\log\rho$ only. Instead of choosing this approximation we will rather attempt to make some statement about the expected manifold from an F--theory perspective. We first embed D7--branes in the non--susy PT setup, neglecting any back--reaction on the internal manifold and then lift the resulting warped resolved conifold with non--trivial axion--dilaton to F--theory. The resulting four--fold is in general not a fibration over a Calabi--Yau three--fold, even in the orientifold limit (see section \ref{ftheory} for this discussion). Solving the full equations of motion would require us to determine the Ricci tensor of the internal manifold from
\begin{equation}\label{zokepe}
  R_{mn} \,=\, \frac{\partial_m\tau\partial_n\bar\tau+\partial_n\tau\partial_m\bar\tau}{4(\rm{Im}\,\tau)^2}
    +\left(T_{mn}^{\rm D7}-\frac{1}{8}\,g_{mn}T^{\rm D7}\right)\,,
\end{equation}
where $T_{mn}^{\rm D7}$ is the energy momentum tensor of the D7 evaluated in our non--trivial background. 
However, we can rely on the fact that in a consistent F-theory compactification this equation is automatically satisfied \cite{GKP} when several stacks of D7-branes and O7-planes are taken into account. An actual computation of the RHS of 
\eqref{zokepe} is generically difficult. This is because to compute $T_{mn}$ of the D7 branes we would first need to
evaluate the non-abelian Born-Infeld action for $N$ D7 branes, and secondly extend the action to curved space because the 
D7 branes wrap non-trivial surfaces in the internal space. We have not been able to perform this direct computation 
(because of the absence of adequate technology), but we give an indirect confirmation of our background from F-theory in
the next section.

Consider first the Bianchi identity, which in leading order becomes ($H_3$ indicates the unmodified NS flux from \eqref{fluxres}, whereas the hat indicates the corrected flux at leading order)
\begin{eqnarray}
  d\hat G_3 & =& d\hat F_3 - d \tau \wedge\hat H_3 - \tau \wedge d\hat H_3 = -d \tau \wedge H_3 +\mathcal{O}((g_s N)^2)  \\ \nonumber
  & = & -\bigg( \frac{N}{2 \pi i } \frac{dz}{z} \bigg) \wedge \big( df_1(\rho) \wedge  d\theta_1 \wedge \sin\theta_1\,d\phi_1 
    + df_2(\rho) \wedge  d\theta_2 \wedge \sin\theta_2\,d\phi_2 \big) +\mathcal{O}((g_s N)^2)\,.
\end{eqnarray}
In order to find a 3--form flux that obeys this Bianchi identity, we make an ansatz
\begin{equation}
  \hat G_3 \,=\, \sum \alpha_i\,\eta_i
\end{equation}
where $\{\eta_i\}$ is a basis of imaginary self--dual (ISD) 3--forms on the resolved conifold. In accordance with the observations about the cohomology of $G_3$, we do not restrict ourselves to (2,1) forms, but allow for $\eta_i$ of (1,2) cohomology as well. With the convention \eqref{cs} we define
\begin{eqnarray}\label{eta}\nonumber
  \eta_1 &=& E_1\wedge E_2\wedge \overline{E}_2 - E_1\wedge E_3\wedge \overline{E}_3\\ \nonumber
  \eta_2 &=& E_1\wedge E_2\wedge \overline{E}_3 - E_1\wedge E_3\wedge \overline{E}_2\\ \nonumber
  \eta_3 &=& E_1\wedge E_2\wedge \overline{E}_1 + E_2\wedge E_3\wedge \overline{E}_3\\ \nonumber
  \eta_4 &=& E_1\wedge E_3\wedge \overline{E}_1 - E_2\wedge E_3\wedge \overline{E}_2\\ 
  \eta_5 &=& E_2\wedge E_3\wedge \overline{E}_1
\end{eqnarray}
\begin{eqnarray} \nonumber
  \eta_6 &=& E_1\wedge \overline{E}_1\wedge \overline{E}_3 + E_2\wedge \overline{E}_2\wedge \overline{E}_3\\ \nonumber
  \eta_7 &=& E_1\wedge \overline{E}_1\wedge \overline{E}_2 - E_3\wedge \overline{E}_2\wedge \overline{E}_3\\ \nonumber
  \eta_8 &=& E_2\wedge \overline{E}_1\wedge \overline{E}_2 + E_3\wedge \overline{E}_1\wedge \overline{E}_3\\ \nonumber
\end{eqnarray}
Note that there are five (2,1) ISD forms, but only three (1,2) ISD forms. This is due to the fact that a form of type (1,2) can only be ISD if it is proportional to $J$.

Not surprisingly, there is no solution to the Bianchi identity involving only the (2,1) forms.
We find a particular solution in terms of only four of above eight 3--forms
\begin{eqnarray}\label{particular}
  P_3 &=& \alpha_1(\rho)\,\eta_1 + e^{-i\psi/2}\alpha_3(\rho,\theta_1)\,\eta_3 + e^{-i\psi/2}\alpha_4(\rho,\theta_2)\,\eta_4 
    +\alpha_8(\rho)\,\eta_8\,,
\end{eqnarray}
with 
\begin{eqnarray}\nonumber\label{alphas}
  \alpha_1 &=& \frac{3g_sNP}{8\pi\rho^3}\frac{ \left[18 a^2 - 36(\rho^2+3a^2)\log\left(\frac{\rho}{a}\right) 
    + (10\rho^2+72a^2)\log\left(\frac{\rho^2}{\rho^2+9a^2}\right)\right]} {\sqrt{\rho^2+6a^2}\sqrt{\rho^2+9a^2}}\\[1ex]
  \nonumber
  \alpha_3 &=& -3\sqrt{6} g_sNP\,\frac{72 a^4-3\rho^4+a^2\rho^2(\log(\rho^2+9 a^2)-56\log \rho)}
    {8\pi\rho^3(\rho^2+6 a^2)^2}\,\cot\frac{\theta_1}{2}\\[1ex]
  \alpha_4 &=& -9\sqrt{6} g_sNP\,\frac{\rho^2-9a^2\log(\rho^2+9 a^2)}{8\pi\rho^4\sqrt{\rho^2+6 a^2}}\,
    \cot\frac{\theta_2}{2}\\[1ex] \nonumber
  \alpha_8 &=& \frac{3a^2}{\rho^2+3a^2}\left[3g_sNP\frac{-9(\rho^2+4a^2)+28\rho^2\log\rho+(81a^2+13\rho^2)\log(\rho^2+9a^2)}
    {8\pi\rho^3\sqrt{\rho^2+6a^2}\sqrt{\rho^2+9a^2}}+\alpha_1\right]\,.
\end{eqnarray}
Note that $a_8$ is implicitly given by $\alpha_1$.
Furthermore, we find a homogeneous solution
\begin{eqnarray}
  G_3^{hom} &=& \beta_1(z,\rho)\,\eta_1 + e^{-i\psi/2}\beta_3(\rho,\theta_1)\,\eta_3 
    + e^{-i\psi/2}\beta_4(\rho,\theta_2)\,\eta_4 \\ \nonumber 
  & &  + e^{-i\psi}\beta_5(\rho,\theta_1,\theta_2)\,\eta_5 +\beta_8(z,\rho)\,\eta_8\,,
\end{eqnarray}
with $\beta_i$ given in \eqref{betas}. This solution has the right singularity structure at $z=0$ and $\rho=0$, but it does not transform correctly under $SL(2,\mathbb{Z})$. When $\psi\to\psi+4\pi$, the axion--dilaton transforms as $\tau\to\tau+N$. This would imply that $G_3$ has to be invariant under this shift, which is true for the particular solution, but not the homogeneous one. We therefore conclude that the correction to the 3--form flux, which is in general a linear combination of $P_3$ and $G_3^{hom}$, is given by \eqref{particular} only
\begin{equation}
  \hat{G}_3 \,=\, G_3+P_3\,.
\end{equation}
Note that in terms of $\eta_i$ the original 3--form flux was given by
\begin{equation}
  G_3 \,=\, -9 P\,\frac{(\rho^2+3a^2)\,\eta_1+ 3a^2\,\eta_8}{\rho^3\sqrt{\rho^2+6a^2}\sqrt{\rho^2+9a^2}}\,.
\end{equation}
We can now determine the change in the remaining fluxes and the warp factor, at least to linear order in $(g_sN)$. 
We find the corrected RR and NS flux from the real and imaginary part of $\hat G_3$, respectively
\begin{equation}
  \hat H_3 \,=\, \frac{ \overline{\hat{G}}_3 - \hat{G}_3 }{\tau - \bar{\tau}}\qquad\mbox{and}\qquad 
    \widetilde{F}_3 \,=\, \frac{\hat{G}_3 + \overline{\hat{G}}_3}{2}\,.
\end{equation}
This results in the closed NS-NS 3--form
\begin{eqnarray}\nonumber
  \hat H_3 &=& d\rho\wedge e_\psi\wedge(c_1\,d\theta_1+c_2\,d\theta_2) + d\rho\wedge(c_3\sin\theta_1\,d\theta_1\wedge d\phi_1-c_4\sin\theta_2\,d\theta_2\wedge d\phi_2)\\
  & & +\left(\frac{\rho^2+6a^2}{2\rho}\,c_1\sin\theta_1\,d\phi_1 
    -\frac{\rho}{2}\,c_2\sin\theta_2\,d\phi_2\right)\wedge d\theta_1\wedge d\theta_2
\end{eqnarray}
and the non--closed RR 3--form (note that $\widetilde{F}_3=\hat F_3-C_0 \hat H_3$, where $\hat F_3$ is closed)
\begin{eqnarray}\nonumber
  \widetilde{F}_3 &=& -\frac{1}{g_s}\,d\rho\wedge e_\psi\wedge(c_1\sin\theta_1\,d\phi_1+c_2\sin\theta_2\,d\phi_2)\\ \nonumber
  & &  +\frac{1}{g_s}\,e_\psi\wedge(c_5\sin\theta_1\,d\theta_1\wedge d\phi_1-c_6\sin\theta_2\,d\theta_2\wedge d\phi_2)\\ 
  & & -\frac{1}{g_s}\,\sin\theta_1\sin\theta_2\left(\frac{\rho}{2}\,c_2 \,d\theta_1-\frac{\rho^2+6a^2}{2\rho}\,c_1\,d\theta_2 \right)
    \wedge d\phi_1\wedge d\phi_2\,,
\end{eqnarray}
see \eqref{defc} for the coefficients $c_i$.
This allows us to write the NS 2--form potential ($dB_2=\hat H_3$)
\begin{eqnarray}\label{bfield}
  B_2 &=& \left(b_1(\rho)\cot\frac{\theta_1}{2}\,d\theta_1+b_2(r)\cot\frac{\theta_2}{2}\,d\theta_2\right)\wedge e_\psi\\ \nonumber
  & + &\left[\frac{3g_s^2NP}{4\pi}\,\left(1+\log(\rho^2+9a^2)\right)\log\left(\sin\frac{\theta_1}{2}\sin\frac{\theta_2}{2}\right)
    +b_3(\rho)\right]\sin\theta_1\,d\theta_1\wedge d\phi_1\\ \nonumber 
  & - & \left[\frac{g_s^2NP}{12\pi\rho^2}\left(-36a^2+9\rho^2+16\rho^2\log\rho+\rho^2\log(\rho^2+9a^2)\right)
    \log\left(\sin\frac{\theta_1}{2}\sin\frac{\theta_2}{2}\right)+b_4(\rho)\right]\\ \nonumber
  & & \qquad\qquad \times \sin\theta_2\,d\theta_2\wedge d\phi_2\,,
\end{eqnarray}
where the coefficients are given in \eqref{defb}. This mirrors closely the result for the singular conifold \cite{ouyang} and we can indeed show that we produce this result in the $a\to 0$ limit. Away from the singular limit, we find an asymmetry between the $(\theta_1,\phi_1)$ and $(\theta_2,\phi_2)$ spheres, which was to be expected since our manifold (the resolved conifold or its more complicated cousin, the resolved warped deformed conifold) does not have the $\mathbb{Z}_2$ symmetry that exchanges the two 2--spheres in the singular conifold geometry. The lesser degree of symmetry is naturally also expressed in the fluxes.

The five--form flux is as usual given by ($\tilde{*}_{10}$ indicates the Hodge star on the full 10--dimensional {\em warped} space)
\begin{equation}
  \hat{F}_5 \,=\, (1+\tilde{*}_{10})(d\hat h^{-1}\wedge d^4x)\,,
\end{equation}
which requires knowledge of the warp factor $\hat h(\rho)$ that is consistent with these new fluxes. In order to solve the supergravity equations of motion one requires
\begin{equation}\label{wfactor}
  \hat h^2\,\Delta \hat h^{-1}-2 \hat h^3\,\partial_m \hat h^{-1}\,\partial_n \hat h^{-1} g^{mn}\,=\,-\Delta\hat{h}
    \,=\,*_6 \left(\frac{\hat G_3\wedge \overline{\hat G}_3}{6(\overline{\tau}-\tau)}\right)\,=\,\frac{1}{6}\,*_6 d\hat{F}_5\,,
\end{equation}
where $\Delta$ is the Laplacian on the unwarped resolved conifold and all indices are raised and lowered with the unwarped metric.
After some simplifications the Laplacian on the resolved conifold takes the form
\begin{equation*}
  \Delta\hat{h} \,=\,\kappa\,\partial_\rho^2 \hat{h} + \frac{5\rho^2+27a^2}{\rho(\rho^2+6a^2)}\,\partial_\rho\hat{h} 
    + \frac{6}{\rho^2}\,\left(\partial_{\theta_1}^2 \hat{h}+\cot\theta_1\,\partial_{\theta_1}\hat{h}\right) 
    + \frac{6}{\rho^2+6a^2}\,\left(\partial_{\theta_2}^2 \hat{h}+\cot\theta_2\,\partial_{\theta_2} \hat{h}
    \right)\,.
\end{equation*}
This should be evaluated in linear order in N, since we solved the SuGra eom for the fluxes also in linear order. As the the right hand side of
\begin{eqnarray}\nonumber
  \frac{1}{6}\,*_6 d\hat{F}_5 &=& 
    \frac{54g_sP}{\pi\rho^6(\rho^2+6a^2)(\rho^2+9a^2)}\Bigg\{12\pi\rho^4+9a^2\rho^2(8\pi-g_sN)+54a^4(4\pi+g_sN)\\ \nonumber
  & & +g_sN\bigg[(25\rho^4+66a^2\rho^2-54a^2)\log\rho+(10\rho^4+102a^2\rho^2+189a^4)\log(\rho^2+9a^2)\\ 
  & & \qquad\quad +6(\rho^4+6a^2\rho^2+18a^4)\log\left(\sin\frac{\theta_1}{2}\sin\frac{\theta_2}{2}\right)\bigg]\Bigg\} 
\end{eqnarray}
appears sufficiently complicated, we need to employ some simplification. The obvious choice is to consider $\rho\gg a$, i.e. we only trust our solution sufficiently far from the tip. As in the limit $a\to 0$ we recover the singular conifold setup, we know our solution takes the form \cite{ouyang}
\begin{eqnarray}
  \hat h(\rho,\theta_1,\theta_2) &=& 1+\frac{L^4}{r^4}\,\left\{1+\frac{24g_sP^2}{\pi\alpha'Q}\,\log\rho\left[1+\frac{3g_sN}{2\pi\alpha'}
    \left(\log\rho+\frac{1}{2}\right)\right.\right.\\ \nonumber
  && \left.\left. \qquad\qquad\qquad\qquad +\frac{g_sN}{2\pi\alpha'}\,\log\left(\sin\frac{\theta_1}{2}\sin\frac{\theta_2}{2}\right)\right]\right\} 
    +\mathcal{O}\left(\frac{a^2}{\rho^2}\right)
\end{eqnarray}
with $L^4=27\pi g_s\alpha'Q/4$. Apart from the $a^2/\rho^2$--correction, this is the same result as for the singular conifold \cite{ouyang}. We have not been able to find an analytic solution at higher order, but considering that most models work with even cruder approximations of the warp factor (i.e. $h(r)\sim\log r/r^4$), we believe this should suffice.

\subsection{D-terms from non--primitive background flux on D7--branes}\label{dterms}

Soft supersymmetry breaking via D--terms on D7--branes has been considered in \cite{uranga}, and was later applied to more realistic type IIB orientifolds \cite{hans, haacklust} or their F--theory lift \cite{lust, berglund} (see also \cite{zwirner} for a IIA scenario); the most general study for generalised CYs has appeared in \cite{luca1}. The established consensus is that non--primitive flux on the D7--worldvolume gives rise to D-terms in the effective 4--dimensional theory, which can only under certain conditions remain non--zero in the vacuum. One way to phrase the necessary condition is to require that the 4--cycle wrapped by the D7--branes admits non--trivial 2--forms that become trivial in the ambient Calabi--Yau, i.e. the $H^2$--cohomology on the four--cycle is bigger than just the pullback of $H^2(CY)$. (Equivalently \cite{hans} states that the 4--cycle needs to intersect its orientifold image over a 2--cycle that supports non--trivial flux. The same is true in the case of two stacks \cite{haacklust} intersecting over a 2--cycle.) This condition can be satisfied for the Ouyang embedding in the $\mu\ne 0$ case: The resolved conifold admits only one non--trivial 2--cycle, the sphere that remains finite at the tip. The 4--cycle that the D7  wraps, on the other hand, can also have a non-trivial cycle spanned by $(\theta_1,\phi_1)$, if the D7 in the Ouyang embedding do not reach all the way to the bottom of the throat. On the D7, this cycle will never shrink completely. Nevertheless, we are mostly concerned with the case $\mu=0$ here. In contrast to \cite{hans, haacklust}
we consider the pullback of a background field with non--vanishing fieldstrength, not the zero mode fluctuations, i.e. we do not expand the worldvolume flux in a basis of $H^2$. This gives rise to a D-term that depends on the overall volume of the manifold and the resolution parameter $a$.  
Though an orientifold will be necessary to consistently compactify our background, we will not specify any orientifold action here, as we do not know a specific compactification our background.

Following the derivation in \cite{haacklust, luca1}, we extract the D-terms from the DBI action. Suppose our stack of N D7--branes wraps a 4-cycle $\Sigma$ as specified by the Ouyang embedding in section \ref{ouyangbed}. The full DBI action for the 8--dimensional worldvolume (in string frame) reads
\begin{equation}\label{d7dbi}
  S_{D7} \,=\, -\mu_7\int_{\Sigma\times\mathcal{M}_4} d^8\xi\,e^{-\Phi}\sqrt{|\check g+\check B-2\pi\alpha'F|}
\end{equation}
where the symbol $\,\check{\phantom{}}\,$ indicates the pullback of the metric and the NS field onto the D7, $F$ is the worldvolume gauge flux. With this product ansatz for the spacetime this expression becomes
\begin{equation}
  S_{D7} \,=\, -\mu_7\int d^4x\,e^{-\Phi}\sqrt{|\check g_4|}\,\sqrt{\left|1+2\pi\alpha'\check g_4^{-1}F_4\right|}\,\Gamma\,,
\end{equation}
where $g_4$ and $F_4$ indicate the 4--dimensional part of the metric and gauge flux and one defines
\begin{equation}\label{defGamma}
  \Gamma \,=\, \int_\Sigma d^4\xi\,\sqrt{|\check g_\Sigma+\mathcal{F}|}\,,
\end{equation}
where we have introduced $\mathcal{F}=\check B-2\pi\alpha'F$. In the following, the pullback is always understood as onto the 4--cycle $\Sigma$. We do not consider any gauge fields along the external space $\mathcal{M}_4$.
The quantity \eqref{defGamma} is the main parameter for the D--terms. Expanding the full action \eqref{d7dbi} at low energies yields the potential contribution
\begin{equation}\label{d7pot}
  V_{D7} \,=\, \mu_7e^{3\Phi}\mathcal{V}^{-2}\Gamma\,,
\end{equation}
where the volume $\mathcal{V}$ of the resolved conifold is defined as
\begin{equation}
  \mathcal{V} \,=\, \frac{1}{6}\int_Y J\wedge J\wedge J\,=\,\frac{(4\pi)^2}{108}\,\int_0^R \rho^3(\rho^2+6a^2)\,d\rho
    \,=\, \frac{8\pi^3}{81}\,R^4(R^2+9 a^2)\,.
\end{equation}
This integral has to be regularised by an explicit cut--off, as we study the non--compact case. Simply cutting off the radial direction does probably destroy the holomorphicity condition, but we will ignore this subtlety here.

One can write \cite{haacklust} $\Gamma=\tilde\Gamma e^{-i\zeta}=|\tilde\Gamma|e^{i(\tilde\zeta-\zeta)}$, where $\zeta$ is determined from the BPS calibration condition and
\begin{equation}\label{tildegamma}
  \tilde\Gamma \,=\, \frac{1}{2}\int_\Sigma \left(\check J\wedge \check J-\mathcal{F}\wedge \mathcal{F}\right)
     +i \int_\Sigma \check J\wedge \mathcal{F}\,.
\end{equation}
Then the condition for the D7 to preserve the same supersymmetry as the O7 corresponds to $\zeta=\tilde\zeta=0$, or equivalently $\Im \tilde\Gamma=0$. Allowing for a small supersymmetry breaking one expands the D7--potential \eqref{d7pot} in $\Im\tilde\Gamma\ll\Re\tilde\Gamma$ and finds
\begin{eqnarray}\nonumber
  V_{D7} &=& \mu_7e^{3\Phi}\mathcal{V}^{-2}\Gamma \,=\, \mu_7e^{3\Phi}\mathcal{V}^{-2}\,
    \sqrt{(\Re\tilde\Gamma)^2+(\Im\tilde\Gamma)^2}\\
  &=& \mu_7e^{3\Phi}\mathcal{V}^{-2}\,\Re\tilde\Gamma + \frac{1}{2}\,\mu_7e^{3\Phi}\mathcal{V}^{-2}\,
    \frac{(\Im\tilde\Gamma)^2}{\Re\tilde\Gamma}\,.
\end{eqnarray}
The first term in this expansion will be cancelled by the tadpole cancellation condition in a consistent compactification. The second term is interperted as the susy--breaking D--term.
The real and imaginary part of $\tilde\Gamma$ are easily read off from \eqref{defGamma} (the integrals are real) and can be calculated for our explicit case at hand. All we need to know is the pullback of the K\"ahler form onto the 4--cycle and the worldvolume flux $\mathcal{F}$.

We would like to consider the simple case such that
\begin{equation}
  \check B \,\ne\,0\,,\qquad\qquad F\,=\,0\,,
\end{equation}
as we have an explicit solution of this form. There could be gauge flux on the D7--brane to could restore supersymmetry in the $a\to 0$ limit. It is noted again that to preserve supersymmetry, holomorphicity is not enough. One also needs the worldvolume flux to be of pure (1,1) type and primitive \cite{gomis}. The reason that it is so difficult to achieve non--trivial D--terms with closed $\check B$ is that $F$ could always cancel the non--primitive part of $\check B$ \cite{hans}, unless some non--trivial topological conditions are met.

In calculating the D--terms, we must treat the D7 as a probe. Thus the B--field that is pulled back is not the one we calculated in \eqref{bfield}, but the original PT solution
\begin{equation}
  B \,=\, f_1(\rho) \sin\theta_1\,d\theta_1\wedge d\phi_1+ f_2(\rho) \sin\theta_2\,d\theta_2\wedge d\phi_2\,,
\end{equation}
where $f_1$ and $f_2$ were defined in \eqref{deffs}. The embedding $z=0$ we use has actually 2 branches, since  
\begin{equation}
   z \,=\, 0 \, =\,  \left ( 9 a^2 \rho^4 + \rho ^6 \right ) ^{1/4} \,\sin\frac{\theta_1}{2}\,\sin\frac{\theta_2}{2}
\end{equation}
can be satisfied by either $\theta_1=0$ or $\theta_2=0$. This implies that also $\phi_1=$fixed or $\phi_2=$fixed, as $\theta_i$ being zero refers to the pole of one of the 2--spheres where the circle described by $\phi_i$ collapses. The full holomorphic cycle is then a sum over these 2 branches.

Consider the 2 four--cycles $\Sigma_1=(\rho,\,\psi,\,\phi_1,\,\theta_1)$ and $\Sigma_2=(\rho,\,\psi,\,\phi_2,\,\theta_2)$ that correspond to the branches $\theta_2=0$ and $\theta_1=0$, respectively. The complex structure induced on them is actually a trivial pullback of the complex structure on the resolved conifold. Using the complex vielbeins \eqref{cs}, we see that 
\begin{equation}
  \Sigma_1 \,=\, (E_1\vert_{\theta_2=0}, E_2)\,,\qquad\qquad \Sigma_2 \,=\, (E_1\vert_{\theta_1=0}, E_3)\,,
\end{equation}
where in $E_1\vert_{\theta_2=0}$ and $E_1\vert_{\theta_1=0}$ the imaginary part is truncated to
\begin{equation}\nonumber
  \Im E_1\vert_{\theta_2=0} \,=\, \frac{\rho\sqrt{\kappa}}{3}\,(d\psi+\cos\theta_1\,d\phi_1)\qquad\mbox{and}\qquad
  \Im E_1\vert_{\theta_1=0} \,=\, \frac{\rho\sqrt{\kappa}}{3}\,(d\psi+\cos\theta_2\,d\phi_2)\,,
\end{equation}
respectively.
It is easy to show that the induced complex structure on the four--cycle still allows for a closed K\"ahler form.
With this observation we find the pullback of $B$ onto both branches
\begin{equation}\label{Bpullback}
  \check B\vert_{\Sigma_1} \,=\, \frac{-3i}{\rho^2}\,f_1\,E_2\wedge \bar E_2\,,\quad\qquad
  \check B\vert_{\Sigma_2} \,=\, \frac{-3i}{\rho^2+6 a^2}\,f_2\,E_3\wedge \bar E_3\,,
\end{equation}
which turn out to be of type (1,1). But that does not mean they are primitive. In fact, as we will see shortly, the pullback of B is not primitive on each individual branch, but in the limit $a\to 0$ the D-term generated by them vanishes when summing over both branches. So it appears that the Ouyang embedding in the singular conifold \cite{ouyang} breaks supersymmetry due to this non--primitivity, but generates neither an F-term nor a D-term. 
Supersymmetry could possibly be restored by choosing appropriate gauge flux,
but we solved the equations of motion only for the case $F=0$, so we will keep working with this assumption. In general, $F$ would mix with the metric in the e.o.m., changing our original setup.

If we consider the B--field \eqref{bfield} that reflects the D7--backreaction, we find its pullback onto $\Sigma_1$ (the case of $\Sigma_2$ is completely analogous)
\begin{eqnarray}\label{fullBpull}
   \check B_2\vert_{\Sigma_1} &=& b_1(\rho)\cot\frac{\theta_1}{2}\,d\theta_1\wedge (d\psi+\cos\theta_1\,d\phi_1)\\ \nonumber
  & + &\left[\frac{3g_s^2NP}{4\pi}\,\left(1+\log(\rho^2+9a^2)\right)\log\left(\sin\frac{\theta_1}{2}\cdot 0\right)
    +b_3(\rho)\right]\sin\theta_1\,d\theta_1\wedge d\phi_1\,.
\end{eqnarray}
We encounter the usual problem that B contains terms with $\log z$, so naturally we find a log--divergent term if we pull back onto a cycle that is described by $z=0$. However, this is not our concern here. We just want to point out, that this B-field is not of pure (1,1) type anymore, but rather contains (2,0) and (0,2) terms as well:
\begin{eqnarray}\nonumber
  \check B_2\vert_{\Sigma_1} &=& \frac{3\sqrt{3}i\,b_1(\rho)}{2\rho^2\sqrt{2\kappa(\rho)}}\,\cot\frac{\theta_1}{2}\left[e^{i\psi/2}
    (E_1\wedge\bar E_2-\bar E_1\wedge \bar E_2)+e^{-i\psi/2}(E_1\wedge E_2+E_2\wedge \bar E_1)\right]\\ 
  &-&\frac{3i}{\rho^2}\left[\frac{3g_s^2NP}{4\pi}\,\left(1+\log(\rho^2+9a^2)\right)\log\left(\sin\frac{\theta_1}{2}\cdot 0\right)
    +b_3(\rho)\right]\,E_2\wedge \bar E_2\,.
\end{eqnarray}
For our considerations the probe approximation shall suffice. We could not obtain any sensible result with the B--field \eqref{fullBpull} anyway, as we would have to integrate over the divergent points $\theta_i=0$. Naturally, this is some kind of self--interaction and divergent. 

Let us now turn to the calculation of the D-terms for the embedding $\mu=0$. The crucial integral for the D-term coming from \eqref{tildegamma} is given by the pullbacks of $J$ and $B$. We still need to give the pullback of $J$ onto both branches:
\begin{eqnarray}\nonumber
  \check J\vert_{\Sigma_1} &=& \frac{\rho}{3}\,d\rho\wedge(d\psi+\cos\theta_1\,d\phi_1)+\frac{\rho^2}{6}\,\sin\theta_1\,d\phi_1\wedge 
    d\theta_1\\
  \check J\vert_{\Sigma_2} &=& \frac{\rho}{3}\,d\rho\wedge(d\psi+\cos\theta_2\,d\phi_2)+\frac{\rho^2+6a^2}{6}\,\sin\theta_2\,d\phi_2\wedge 
    d\theta_2\,.
\end{eqnarray}
And we repeat the pull--back of $B$ in terms of real coordinates:
\begin{equation}\label{Bpullreal}
  \check B\vert_{\Sigma_1} \,=\, f_1(\rho)\,\sin\theta_1\,d\theta_1\wedge d\phi_1\,,\quad\qquad
  \check B\vert_{\Sigma_2} \,=\, f_2(\rho)\,\sin\theta_2\,d\theta_2\wedge d\phi_2\,.
\end{equation}
The D-term is now obtained from $\Im\tilde\Gamma$ in \eqref{tildegamma}
\begin{eqnarray}\nonumber\label{dterm}
  D &=& \int_{\Sigma_1} \check J\vert_{\Sigma_1}\wedge \check B\vert_{\Sigma_1}+ \int_{\Sigma_2} \check J\vert_{\Sigma_2}\wedge \check 
    B\vert_{\Sigma_2}\\
  &=& \int_{\Sigma_1} \frac{\rho}{3}\,f_1\sin\theta_1\,d\rho\wedge d\psi\wedge d\theta_1\wedge \phi_1
    + \int_{\Sigma_2} \frac{\rho}{3}\,f_2\sin\theta_2\,d\rho\wedge d\psi\wedge d\theta_2\wedge \phi_2\,.
\end{eqnarray}
We see immediately that for the case $f_1=-f_2$, i.e. the singular $a\to 0$ limit of the KT solution, the D-term vanishes after summing both cycles, even though the pullback of $B$ is non-primitive in this case.
For the case $a\ne 0$ we can perform the integrals, again introducing a cut--off $R$ for the radial direction. We find
\begin{equation}\label{dresult}
  D \,=\, \frac{32\pi^2g_sP}{9}\,\big[9a^2\log (9+a^2)-(9a^2-2R^2)\log R-(9a^2+R^2)\log(9a^2+R^2)\big]\,.
\end{equation}

To obtain the full D-term potential, we also need $\Re\tilde\Gamma$ from \eqref{tildegamma}. Looking at the pullbacks of the B--fields \eqref{Bpullback} we see that $\check B\wedge \check B$ vanishes for both branches, so
\begin{eqnarray}\nonumber
  \Re\tilde\Gamma &=& \frac{1}{2}\int_{\Sigma_1} \check J\vert_{\Sigma_1}\wedge  \check J\vert_{\Sigma_1} + \frac{1}{2}\int_{\Sigma_2} 
    \check J\vert_{\Sigma_2} \wedge  \check J\vert_{\Sigma_2}\\ 
  &=& \frac{4\pi^2}{9}\,R^2(R^2+6a^2)\,.
\end{eqnarray}
The total D-term potential then reads
\begin{eqnarray}\nonumber
  V_{D7} &=& \frac{1}{2}\,\mu_7e^{3\Phi}\mathcal{V}^{-2}\,\frac{(\Im\tilde\Gamma)^2}{\Re\tilde\Gamma}\\
  &=& \frac{59049\,\mu_7\,e^{3\Phi}}{512 \pi^8}\,\frac{D^2}{R^{10}(R^2+6a^2)(R^2+9a^2)^2}
\end{eqnarray}
with the D-term $D$ from \eqref{dresult}. In the probe approximation, $\Phi$ is just the constant background dilaton and can be set to zero. This is one of the main results of our paper. We find a non--zero D--term created by non-primitive (1,2) flux when pulled back to non-primitive flux on D7--branes. Their magnitude is highly suppressed in a large volume compactification. It would be most desireable to find a consistent compactification for our setup, in which we do not have to introduce a cut--off by hand that spoils holomorphicity. Let us stress again that these (1,2) fluxes did not lead to the creation of a bulk cosmological constant, because they are ISD. We would expect, however, a modification of the superpotential, i.e. in general D-terms on D7--branes also create F-terms \cite{lust, hans, haacklust}. 

We have so far neglected  any zero modes. Once we study D3/D7 inflation, there will also be degrees of freedom that become light when the two branes approach each other. The D-- and F--terms in this case have to be re-evaluated. As already outlined in the beginning of this section, we believe that the conditions to have non--zero D--terms in the vacuum (i.e. intersection over a two--cycle with non--trivial flux or a cohomology $H^2(\Sigma)$ of the 4--cycle that is greater than the pullback of the CY cohomology $H^2(CY)$) can be met when $\mu\ne 0$. For $\mu=0$ it appears rather the opposite:
There is only one non--trivial 2--cycle in the resolved conifold, the blown--up $(\phi_2,\,\theta_2)$--sphere. With $\mu=0$, the cycle $\Sigma_1$ is topologically trivial (it contains the shrinking 2--sphere), the cycle $\Sigma_2$ is not. However, once we compactify we will introduce another cycle on which the (0,1) form is supported. This should be in $(\rho,\,\psi)$ direction, as $G^{(1,2)}\sim J\wedge \bar E_1$, and $E_1$ extends along $\rho$ and $\psi$. However, from \eqref{Bpullreal} we see that this 2--cycle does not support any flux.

We believe this puzzle might be clarified once the original Ouyang embedding in the singular conifold background is made supersymmetric with appropriate gauge fluxes. Note however, that there is an essential difference between the singular KT and the resolved PT backgrounds: the B--field in the bulk is primitive, i.e. $J\wedge J\wedge B=0$, for the first case but not for the latter.

The next step would be to consider the embedding $\mu\ne 0$. The integrals becomes much more complicated and cannot be solved analytically. Only for the case $a=0$ have we been able to show by numerical integration that $D=0$. For $a\ne 0$ the integrand's strong oscillatory behaviour has prevented us from finding a solution so far. Note that the pullback of $J$ and $B$ is much more involved. We have to use the embedding equations
\begin{equation}\label{rhopsi}
  (\rho^6+9a^2\rho^4) \,=\, \left(\frac{|\mu^2|}{\sin\frac{\theta_1}{2}\,\sin\frac{\theta_2}{2}}\right)^4\,,\quad
  \psi \,=\, \phi_1+\phi_2+\,\mbox{const}\,.
\end{equation}
It is then tedious but straightforward to calculate the pullback
\begin{equation}
  \check J_{\alpha\beta} \,=\, \partial_\alpha y^m \partial_\beta y^n\,J_{mn}\,, 
\end{equation}
where $m,n=\rho,\psi,\theta_1,\theta_2,\phi_1,\phi_2$ run over the whole 3--fold, whereas $\alpha,\beta=\theta_1,\theta_2,\phi_1,\phi_2$ parameterise the 4--cycle. A similar formula gives the pullback of the NS field $\check B$. Note, however, that the pullback will contain terms with $(\sin\theta_i)^{-1}$, which diverge at the integration boundaries $\theta_i=0$. For the case $a=0$ this seems to be under control, for the resolved case we cannot make any definite statement.

\section{A view from F--theory}\label{ftheory}
\setcounter{equation}{0}

Now that we have more or less the complete type IIB picture, 
we should deviate to address the F-theory \cite{Vafa}
lift of our background. Studying F-theory lift has
many advantages:

\vskip.1in

\noindent $\bullet$ It can give us a precise way to study the compact version of our background. 
Recall that the background 
that we constructed is non-compact. The compact form of our background can be formulated if we can find a compact 
four-fold associated with the resolved conifold background.

\noindent $\bullet$  It is directly related to M-theory by a $S^1$ reduction \cite{Vafa}. 
In M-theory the structure of the four-fold
remains the same, but there are a few advantages. 
We can determine the precise warped form of the metric \cite{BeckerM, DRS}, the precise
superpotential \cite{GVW} and the complete perturbative \cite{DRRS}
and non-perturbative terms on the IIB seven branes. 

\vskip.1in

\subsection{Construction of the fourfold}

\vskip.1in

With the above advantages in mind, we aim to determine the fourfold in F-theory and study the subsequent properties associated with the 
fourfold in M-theory. The generic structure of the fourfold can be of the following form:
\begin{equation}
\label{mfourfold}
ds^2 = e^{2A} \eta_{\mu\nu} dx^\mu dx^\nu + e^{2B} g_{mn} dy^m dy^n + e^{2C} \vert dz\vert^2
\end{equation} 
where $A, B, C$ are the warp factors that could be in general functions of time as well as the internal 
coordinates ($y^m, z$) and ($\mu, \nu$) = (0, 1, 2). 
The fourfold is a $T^2$ fibration over a base. We denote the complex coordinate of the 
$T^2$ by $z$ and the base has a metric $g_{mn}$. The corresponding type IIB metric is expected to be of the 
form (see also \cite{chen}):
\begin{equation}
ds^2 ~= ~e^{2A+C} ~(-dx_0^2 + dx_1^2 + dx_2^2)~ +~ 
\frac{e^{-3C}}{\vert\tau\vert^2}~dx_3^2 ~+~
e^{2B+C} ~g_{mn} dy^m dy^n
\end{equation}
which tells us that in principle the $3+1$ dimensional Lorentz could be broken by 
choosing a generic warp factor of the fibre torus in M-theory. The fibre torus, in M-theory, is parametrised by 
a complex structure $\tau$ which is proportional to the axio-dilaton in type IIB:
\begin{equation}
dz ~=~ dx^{11} + \tau dx^{3}, ~~~~~~~ {\tau} ~ = ~ C_0 ~ + ~ \frac{i}{g_s}
\end{equation}
Clearly if the torus was non-trivially fibred over the threefold base (with metric $g_{mn}$) we would expect non-zero 
cross terms in the type IIB metric. For our case we simply choose a trivial $T^2$ fibration of the 
fourfold, so the cross-terms are absent. For a compact manifold we would require the axion charge to vanish. This 
would mean that the contribution to $C_0$ from a single D7 brane is very small. This would change our metric 
to 
\begin{equation}
ds^2 ~= ~e^{2A+C} ~(-dx_0^2 + dx_1^2 + dx_2^2)~ +~ 
\frac{e^{-3C}}{({\rm Im}~\tau)^2}~dx_3^2 ~+~
e^{2B+C} ~g_{mn} dy^m dy^n
\end{equation} 
Furthermore, restoring full $3+1$ dimensional Lorentz 
invariance will tell us that the type IIB 
metric has the following form:
\begin{equation} 
ds^2 ~ = ~ \frac{e^{3A/2}}{\sqrt{{\rm Im}~\tau}} ~ \eta_{\mu\nu}dx^\mu dx^\mu ~ + ~ 
\frac{e^{2B - A/2}}{\sqrt{{\rm Im}~\tau}} ~g_{mn} dy^m dy^n
\end{equation}
Comparing the above form of the metric with the metric that we have \eqref{pzmet}, it is easy to work out the 
corresponding M-theory warp factors in terms of $h$ and the axio-dilaton $\tau$ as:
\begin{equation}
\label{warpy}
e^A ~ = ~ \Bigg[\frac{{\rm Im}~\tau}{h}\Bigg]^{\frac{1}{3}}, ~~~~~ e^B ~ = ~ 
\Big[h ({\rm Im}~\tau)^2\Big]^{\frac{1}{6}}, ~~~~~ e^C ~ = ~  
\Bigg[\frac{\sqrt{h}}{({\rm Im}~\tau)^2}\Bigg]^{\frac{1}{3}}
\end{equation}
Now combining \eqref{warpy} with \eqref{mfourfold} we can easily see that the fourfold is a given by the following 
metric:
\begin{eqnarray}\label{fourfold}\nonumber
  ds^2_{\rm 4-fold} & = & \frac{h^{1/3}}{({\rm Im}~\tau)^{4/3}}~ 
\big\vert dx^{11} + \tau~dx^{3}\big\vert^2 + 
{h^{1/3}({\rm Im}~\tau)^{2/3}}\Big[ 
\frac{d\rho^2}{\kappa} ~ + \frac{\rho^2}{6}\,\big(d\theta_1^2+\sin^2\theta_1\,d\phi_1^2\big) ~ + \\
& & + ~ \frac{\kappa}{9}\,\rho^2\big(d\psi+\cos\theta_1\,d\phi_1
 +\cos\theta_2\,d\phi_2\big)^2 
    +\frac{\rho^2+6a^2}{6}\,\big(d\theta_2^2+\sin^2\theta_2\,d\phi_2^2\big)\Big]\,,
\end{eqnarray}
where the other variables have already been defined above. The type IIB NS and RR three-form fluxes would converge 
to give us $G$-fluxes $G_{mnpq}$ on the fourfold. The equations of motion of $G$-fluxes are determined from the 
gravitational quantum corrections in M-theory as well as $M2$ brane sources. To analyse this on the fourfold 
background \eqref{fourfold} becomes too cumbersome, so let us simply illustrate the case of a metric \eqref{mfourfold} with a warp factor 
of the fibre torus $e^{2B}$ i.e $C = B$. In this case the $G$-fluxes satisfy the following two equations:
\begin{eqnarray}\label{gfluxe}\nonumber
&&(1)~ D_q\Big[e^{3A}\big(G^{mnpq} ~ - ~ (\ast G)^{mnpq}\big)\Big] ~ = ~ \frac{2k^2 T_2}{8!}\epsilon^{mnpa_1....a_8}
(X_8)_{a_1....a_8} \\
&& (2)~ \square ~e^{6B} ~ = ~ -\frac{1}{2\cdot 4!} G_{mnpq}(\ast G)^{mnpq} ~ - ~ 
\frac{2k^2 T_2}{8!}\cdot \frac{X_8}{\sqrt{-g}} + ...
\end{eqnarray}
where $k^2 T_2$ are constants appearing in the M-theory Lagrangian, and we have made 
all fields and the Hodge star operations 
w.r.t. the unwarped metric, except for the $X_8$ term.  
The $X_8$ term in the above two equations is the eight form expressed entirely in terms of 
the curvature tensor of the warped metric. This is the quantum correction that we can put to zero when the background 
is non-compact. A simple observation of \eqref{gfluxe} will tell us that for a compact manifold, a vanishing $X_8$ term 
will lead to contradiction. 

We have also left some dotted terms in the second equation of \eqref{gfluxe}. These unwritten terms account for sources, like $M2$
branes, in the theory. These $M2$ branes are precisely the D3 branes that we will need to eventually put in to study inflation in
our model. 

Observe now that when we make $X_8$ negligibly small (or in other words, when we ignore quantum corrections), 
the equations of motion of the $G$-fluxes \eqref{gfluxe}, tell us that the covariant derivatives of $G$-fluxes have to vanish. This condition can be satisfied by two different varieties of $G$-flux:
\begin{equation}\label{susycond}
G_{mnpq} ~ = ~ (\ast G)_{mnpq}, ~~~~~{\rm or} ~~~~~ G_{mnpq} ~ - ~ (\ast G)_{mnpq} ~ = ~ e^{-3A}\gamma_{mnpq}
\end{equation}
where $\gamma_{mnpq}$ is a covariantly constant tensor. 
The first condition means that the $G$-fluxes have to be self-dual. If it is also 
primitive then this is the condition to 
preserve susy \cite{BeckerM}\footnote{Recall that primitivity implies self-duality but not vice-versa on a 4-fold, in contrast to primitivity and \emph{imaginary} self--duality on a 3--fold.}. The second condition concerns us here. Generically, this implies that
the $G$-fluxes are not primitive and therefore 
susy is spontaneously broken in our model. However, if we can rewrite $\gamma_{mnpq}$ as 
\begin{equation}\label{covgama}
\gamma_{mnpq} ~\equiv~ e^{3a}\gamma^{(1)}_{mnpq} ~ - ~ e^{3a}\left[\ast\gamma^{(1)}\right]_{mnpq}
\end{equation}
with $e^{3a}$ being a function that we will specify below, 
then self-duality is restored in the presence of a new $G$-flux that is of the form 
\begin{equation}\label{newgg}
{\cal G}_{mnpq} \equiv G_{mnpq} - e^{-3(A-a)}\gamma^{(1)}_{mnpq}
\end{equation}
although this may not be primitive.
Indeed, if we demand $\gamma^{(1)}$ to be of the form
\begin{equation}\label{jwedj}
\gamma^{(1)} ~\equiv ~{\cal J} \wedge {\cal J}
\end{equation}
with ${\cal J}$ being the fundamental 2--form in M/F-theory and $e^{-3(A - a)}$ 
is a closed zero form then susy can be broken with a 
non-primitive self-dual (2, 2) form \cite{beckersusy}\footnote{A non-self-dual flux of the form 
$G_{mnpq} = \frac{e^{-3A}}{2}\left(\gamma - \ast \gamma\right)_{mnpq}$ can also break susy and satisfy the 
second condition in \eqref{susycond}. However, such a choice of flux does not satisfy the equation of motion.}. 
A similar condition can be derived on the fourfold with three warp factors, as in \eqref{mfourfold} and \eqref{fourfold}. With three warp factors the analysis remains the same. One 
can easily verify this from the $G$-fluxes constructed out of type IIB three-forms. In the following we will try to 
justify the existence of this (2, 2) non-primitive form. 

\vskip.1in

\subsection{Normalisable harmonic forms and seven branes}

\vskip.1in

So far, our study in M-theory has followed in parallel to that in type IIB. To see some novelty 
from the M-theory picture, let us look for the remnants of the seven branes in M-theory. Since M-theory 
does not support any branes other than two and five-branes, the information of type IIB seven branes can only come from
the gravity solution. In type IIB theory, recall that the seven branes were embedded via the Ouyang embedding \cite{ouyang}.
This means the embedding equation is:
\begin{equation}\label{oembed}
(\rho^6 + 9a^2 \rho^4)^{1/4} ~{\rm exp} \bigg[\frac{i(\psi - \phi_1 - \phi_2)}{2}\bigg]~{\rm sin}~\frac{\theta_1}{2} 
~  {\rm sin}~ \frac{\theta_2}{2} = \mu^2
\end{equation}
In the limit $\mu \to 0$ the seven branes should be embedded via the two branches:
\begin{eqnarray}\label{sevencoor}\nonumber
&&{\rm Branch ~1}:~~~\theta_1 ~ = ~ 0, ~~~~ \phi_1 ~ = ~ 0 \\ 
&&{\rm Branch ~2}:~~~\theta_2 ~ = ~ 0, ~~~~ \phi_2 ~ = ~ 0 
\end{eqnarray}
and both run along the radial direction\footnote{It is easy to see why. 
A generic configuration of seven branes would be able to lower their actions by going to 
smaller $\rho$. Therefore, they cannot be fixed at a specific $\rho \equiv \rho_0$.}. 
The full geometrical analysis of the embedding is difficult, but we can see that for branch 1
 the seven branes wrap a four-cycle along directions ($\theta_2, \phi_2$) and ($\psi, \rho$) inside
the resolved conifold background and are stretched along the spacetime directions $x^{0, 1, 2, 3}$. One can easily see
that the axionic charges of the seven branes could all globally cancel by allowing a trivial F-theory monodromy
so that there is no contradiction with
Gauss' law. Subtleties come when we want to study compact manifolds in the presence of seven-branes {\it and} 
non-primitive fluxes. In the absence of non-primitive fluxes one can compactify the manifold with a sufficient number 
of seven branes and orientifold planes. The more subtle situation with non-primitive fluxes will 
be discussed later.

For the present case let us look at the metric along directions orthogonal to the type IIB seven branes. The M-theory
metric given above \eqref{fourfold} will immediately tell us the orthogonal space to be:
\begin{eqnarray}\label{orthospace}\nonumber
ds^2 ~ & = &~ \frac{h^{1/3}}{({\rm Im}~\tau)^{4/3}} ~\big\vert dx^{11} + \tau dx^{3}\big\vert^2 ~ + ~ 
{h^{1/3} ({\rm Im}~\tau)^{2/3}} \Big[\frac{\rho^2}{6}\,
d\theta_1^2 + \frac{\rho^2}{6} {\rm sin}^2 \theta_1~d\phi_1^2 \Big] \\ 
&= & \frac{h^{1/3}}{({\rm Im}~\tau)^{4/3}} ~(dx^{11} + {\rm Re}~\tau ~dx^{3})^2 + 
{h^{1/3} ({\rm Im}~\tau)^{2/3}} \Big[\frac{\rho^2}{6}d\theta_1^2 + \frac{\rho^2}{6} {\rm sin}^2 \theta_1 ~d\phi_1^2\Big]
 \\ \nonumber
&& ~+ ~\frac{h^{1/3}}{({\rm Im}~\tau)^{4/3}} ~ ({\rm Im}~\tau)^2 (dx^{3})^2
\end{eqnarray} 
where Re $\tau$ and Im $\tau$ are related to the axion and dilaton respectively in the following way:
\begin{eqnarray}\label{axiodil}\nonumber
&&{\rm Re}~\tau ~ \equiv ~ C_0 ~ = ~ \frac{N}{2\pi} (\psi - \phi_1 - \phi_2) \\
&&{\rm Im}~\tau ~ \equiv ~ e^{-\Phi} ~ = ~ \frac{1}{g_s} - \frac{N}{2\pi}~ 
{\rm log}\Bigg[(\rho^6 + 9a^2 \rho^4)^{\frac{1}{4}} ~{\rm sin}~
\frac{\theta_1}{2}~{\rm sin}~\frac{\theta_2}{2}\Bigg]
\end{eqnarray}
and $N$ is the number of the seven branes, as discussed in \cite{ouyang}. The above choice of 
axion-dilaton is {\it not} the full story, as we will discuss in details in the sequel.
 For the time being, however, we 
will continue using this result because the corrections to axion-dilaton are subleading. Some aspects 
of these corrections have been discussed in \cite{BCDF} using results of \cite{DKS}.

To study the geometry further, let us analyse the background close to the point ($\phi_1=0,$ $\theta_1=0$). 
The resulting metric in the local
neighbourhood of the point ($\phi_1, \theta_1$) has the following form:
\begin{equation}\label{locmeti}\nonumber
ds^2 ~=  {h^{1/3} ({\rm Im}~\tau)^{2/3}} \Bigg[\frac{\rho^2}{6} d\theta_1^2 + 
\frac{\rho^2}{6} {\rm sin}^2\theta_1~d\phi_1^2 + (dx^{3})^2\Bigg] + 
\frac{h^{1/3}}{({\rm Im}~\tau)^{4/3}} ~\Bigg(dx^{11} + 
\frac{N}{2\pi} (\psi - \phi_1 - \phi_2) dx^{3}\Bigg)^2
\end{equation}
which can be compared to a Taub-NUT metric:
\begin{equation}\label{Tn}
ds^2_{\rm Taub-NUT} = V({\tilde r}) \Big(dx^{11} + A_{3} dx^{3}\Big)^2 + V({\tilde r})^{-1} 
\Big[ d{\tilde r}^2 + {\tilde r}^2 d\theta^2 + 
{\tilde r}^2 ~{\rm sin}^2~\theta (dx^{3})^2\Big]
\end{equation}
with $V({\tilde r})$ being the typical harmonic function. We see that \eqref{locmeti} does have a strong resemblance to 
\eqref{Tn}, with the $A_{3}$ charge of the Taub-NUT being given by the axionic charge of $N$ type IIB seven-branes,
as expected. However, the local metric is more complicated than the standard TN space because of the non-trivial 
back-reaction of the G-flux. In particular, the warp factors and some of the coordinates 
appearing in \eqref{locmeti} are 
not quite of the form in \eqref{Tn}. Nevertheless, \eqref{locmeti} does capture some of the key features of a 
Taub-NUT space, namely, the $U(1)$ fibration structure and the gauge charge. In 
\eqref{Tn} the gauge charge has a proportionality $A_{3} \propto {\rm cos}~\theta$. Such a choice of Taub-NUT charge helps us to determine
an anti-self-dual harmonic form in this space \cite{gibbsone, senone, DRRS}. Comparing this to \eqref{locmeti}, 
we see that the charge is given by $C_0 \equiv \frac{N}{2\pi}\left(\psi - \phi_1 - \phi_2\right)$. 
A small change in this charge can be related to a small change in $\phi_1$, keeping other variables constant (recall 
that we are measuring the charge away from the D7 brane). 

We now define the vielbeins in the following way:
\begin{eqnarray}\label{vieldef} \nonumber 
&& e^y \equiv \frac{h^{1/6}}{({\rm Im}~\tau)^{2/3}}\Bigg(dx^{11} + 
\frac{N}{2\pi} (\psi - \phi_1 - \phi_2) dx^{3}\Bigg), ~~~~~ e^3 \equiv h^{1/6} ({\rm Im}~\tau)^{1/3}~dx^3 \\
&& e^{\theta_1}\equiv \frac{h^{1/6} ({\rm Im}~\tau)^{1/3} \rho}{\sqrt{6}} ~d\theta_1, ~~~~~~~~~~  
e^{\phi_1}\equiv \frac{h^{1/6}({\rm Im}~\tau)^{1/3} \rho ~{\rm sin}~\theta_1}{\sqrt{6}} ~d\phi_1  \, .
\end{eqnarray}
Using these vielbeins we are now ready to construct our harmonic forms.
These harmonic forms have to be 
self-dual (or anti self-dual)
as well as normalisable. Let us make the following ans\"atze for the one-form:
\begin{equation}\label{hform}
\omega ~ = ~ l({\theta_1}) ~\Bigg(dx^{11} + \frac{N}{2\pi} (\psi - \phi_1 - \phi_2) dx^{3}\Bigg) \, .
\end{equation}
The harmonic two-form will then be given by $d\omega$ and is therefore exact as well as harmonic. To require this to 
be anti self-dual, we want $\ast d\omega = - d\omega$ in this space with the Hodge star 
being given by the warped metric \eqref{locmeti}. This gives us:
\begin{equation}\label{ltheta}
l(\theta_1) ~ = ~ {\rm exp}~\Bigg[\mp \frac{N}{2\pi} \int^{\theta_1} \frac{d\theta_1}{{\rm sin}~\theta_1 ~
{\rm Im}~\tau}\Bigg] \, .
\end{equation}
This implies that the one-form is:
\begin{equation}\label{oneform}
\omega ~ = ~ {\rm exp} ~\Bigg[\mp\int^{\theta_1} \frac{d\theta_1}{{\rm sin}~\theta_1 
\left({\rm log~ sin}~\frac{\theta_1}{2} + ...\right)}\Bigg]
\Bigg(dx^{11} + \frac{N}{2\pi} 
(\psi - \phi_1 - \phi_2) dx^{3}\Bigg) \, ,
\end{equation}
which clearly means that an anti self-dual two-form is normalisable, whereas a 
self-dual two-form in not. 
Existence of such normalisable forms guarantees many things:
firstly it confirms the existence of seven branes in this background. Once the harmonic forms are defined over a 
compact two-sphere then the resulting background can be compactified so that an effective four-dimensional theory 
could be defined. In the presence of a non-compact background, the harmonic forms are very useful to determine 
the world volume theory on the seven branes \cite{imamura, sentwo, DRRS}. Secondly, existence of harmonic forms
guarantees the {\it non-commutative} deformations on the seven-branes \cite{DRRS}. Recall that the world-volume
theory on the type IIB seven-branes is non-commutative because of the presence of non-primitive fluxes. This is 
perfectly consistent with the original D3/D7 inflationary model \cite{DHHK} that was also non-commutative 
due to the presence of a non-primitive background. The key difference between our present background and the original
D3/D7 system is that (apart from being the fact that the original D3/D7 system was defined on $K3 \times T^2/Z_2$)   
in the original D3/D7 system the non-primitivity was treated as a tunable 
parameter (although it might violate the equations of motion) and could be switched off to regain supersymmetry. In our present scenario we see no way to switch off the non-primitivity. In other words, our present background 
is inherently non-supersymmetric.  

At this point we wish to make several comments: Firstly,
the above analysis is only for one of the embedding branches. It is not difficult to see that a similar 
analysis could be performed for the other branch. The total normalisable anti-selfdual harmonic form is 
presumably a linear combination of these two forms.  
Secondly, $-$ and this is important $-$ 
the above analysis relies heavily on the particular embedding that we took, namely the embedding \eqref{oembed}. This
embedding is the trivial embedding that should be modified when $\mu \ne 0$ in \eqref{oembed}. An immediate 
modification of the embedding equation \eqref{sevencoor}, which was for $\mu = 0$, will be the following set 
of equations:
\begin{equation}\label{corembed}
(\rho^6 + 9a^2 \rho^4)^{1/4}~{\rm sin}~\frac{\theta_1}{2}~{\rm sin}~\frac{\theta_2}{2} ~ = ~ \vert \mu\vert^2, ~~~~~~~ 
\psi - \phi_1 - \phi_2~ = ~ \tilde\theta
\end{equation}
where 
$\tilde\theta \equiv -i~{\rm log} ~\frac{\mu}{\vert\mu\vert} - 2n\pi$ is a phase factor 
fixed by the orientation of the seven branes in the angular directions. As soon as 
$\mu \ne 0$, the embedding equations are no longer the simplified equation \eqref{sevencoor}, but rather the surface 
\eqref{corembed}. Thus we see in a resolved conifold that the seven branes wrap along a nontrivial curved four-cycle
in the internal space\footnote{This is clearly a four-cycle because there are six unknowns and two equations in 
\eqref{corembed}.}. 

For this case one can also work out the normalisable harmonic form. The analysis is more complicated but can be worked 
out as before. We will not attempt this here, but end this part of the discussion by noting that these normalisable 
harmonic forms would give rise to second cohomologies (i.e the second Betti numbers) once we compactify the 
non-compact resolved conifold background. 

\vskip.1in

\subsection{One forms and M-theory uplift of fluxes}\label{lift}

\vskip.1in

At this point we should come back to the issue that we briefly 
alluded to earlier: compactifying our manifold in type IIB theory. From the F/M-theory point of view, this is equivalent to finding a consistent compact base. This problem has already been solved earlier in \cite{gtpaper1, gtpaper2} and 
\cite{dot1, dot2, dot3}. The compact base $-$ which we call $B$ henceforth $-$ should have at least one smooth 
curve ${\bf P}^1$ with normal bundle ${\cal O}(-1)\oplus {\cal O}(-1)$. The Weierstrass model for the fourfold 
can be obtained as a Calabi-Yau hypersurface with the equation:
\begin{equation}\label{weierfor}
y^2 ~ = ~ 4 x^3 -g_2 x - g_3
\end{equation}
where $y$ is the coordinate on the bundle ${\cal O}_B(3K_B)$, $x$ is the coordinate on the bundle ${\cal O}_B(2K_B)$  
and $g_k$ is a section of ${\cal O}_B(-2kK_B)$ for $k = 2, 3$. 

The elliptic fibre is generically smooth, but is a cuspidal cubic over points where $y^2 = 4 x^3$ and nodal cubic over 
points where $g_2^3 = 27 g_3^2$ with $g_k$ not zero. These latter are, of course, the points where the discriminant of the 
Weierstrass equation vanishes. The zero locus of the discriminant is a complex surface $S$ containing the curve 
$D$ defined by $y^2 = 4x^3$. Once we know $S$ and $D$, the Euler characteristics of the fourfold can be completely 
written in terms of the Euler characteristics of these submanifolds, i.e
\begin{equation}\label{euler}
\chi ~ = ~ \chi(S) + \chi(D) ~ = ~ 19728 ~ = ~ 24 \times 822
\end{equation}
which would tell us that the total number of branes and fluxes should add up to 822 for this manifold\footnote{
Incidentally, if we make a conifold transition to the base to go to a fourfold that is a $T^2$ fibration over a
deformed conifold base, the Euler number remains unchanged. See \cite{gtpaper2, dot3} for more details.}. 

Observe, however, that the fourfold that we choose with a K\"ahler base is not the most generic answer. In general, the 
base could be a non-K\"ahler manifold. What we need from our present analysis is the existence of one-forms in our 
manifold which could be used to express the (1,2) fluxes in the type IIB set-up. Presently, in the type IIB 
set-up, we can think of the following three choices of one-forms in our manifold:

The first of the three one forms can be written in terms of the holomorphic coordinates ($z, y, u, v$) given in \eqref{holocoord}, in the following way \cite{ionel}:
\begin{equation}\label{oneforma}
\omega_1 \equiv r^{-2}\left(N^{1/3} + 4a^4 N^{-1/3} - 2a^2\right)~{\rm Im}~\left(\bar z dz + \bar y dy  + 
\bar u du  + \bar v dv \right)
\end{equation}
where $N = N(r) = \frac{1}{2}\left(r^4 - 16a^6 + \sqrt{r^8 - 32 a^6 r^4}\right)$. See \eqref{rhoandr} for the relation between $r$ and our radial coordinate $\rho$. The above one form 
contributes an exact part to the K\"ahler form on the resolved conifold. This one form is invariant under the 
underlying $SO(4, \mathbb{R})$. 

Another one form can be constructed using the homogeneous coordinates $\zeta_+ = \frac{\xi_2}{\xi_1}$ and 
$\zeta_- = \frac{\xi_1}{\xi_2}$ that respectively define the two patches $H_+$ where $\xi_1 \ne 0$ and 
$H_-$ where $\xi_2 \ne 0$ on the $S^2$ of the resolved conifold. (See appendix \ref{rescone} for more details on the geometry.) 

We construct one forms on the two patches $H_\pm$ in the following way:
\begin{equation}\label{oneformb}
\omega_\pm ~ = ~ \frac{1}{2}~ {\rm Im}~ \frac{\zeta_\pm d\bar\zeta_\pm}{1 + \vert\zeta_\pm\vert^2}
\end{equation}
One can also show that these forms are also invariant under $SO(4)$ just like $\omega_1$ above. 

Finally, the third category of one forms in our background are of the form:
\begin{equation}\label{oneformc}
\omega_3^i ~ = ~ g_i(\rho) E_i, ~~~~~ \bar\omega_3^j ~ = ~ h_j(\rho) \bar E_j
\end{equation}
with no sum over $i, j$ (although one can combine these one forms to write another one form). The $E_i$ are 
the complex vielbeins described in Appendix A. These one forms can only exist on the compactified base if they have 
a compact support. In the following we will discuss the asymptotic behaviours of $\omega_3$ and $\bar\omega_3$. 

To study the asymptotic behaviour it is important to divide our type IIB fluxes into ($2,1$) and ($1, 2$) parts. Let 
us also scale the radial coordinate $\rho$ as $\rho \to \lambda \rho$ so that large $\lambda$ means that we are 
exploring UV geometries. In this limit clearly 
\begin{equation}\label{vielasym}
E_i ~ \to ~ \lambda E_i, ~~~~~~~~~ \eta_i ~ \to ~ \lambda^3 \eta_i
\end{equation}
where the ISD forms $\eta_i$ were defined in \eqref{eta}. The ($2, 1$) part of 
${\hat G}_3$ is then\footnote{Recall that we are using hatted quantities to indicate the background flux with 
backreaction from the embedded seven branes.}: 
\begin{equation}\label{21part}
{\hat G}_3^{(2, 1)} = \Bigg[\alpha_1(\rho) ~- 
~\frac{9P(\rho^2+3a^2)}{\rho^3\sqrt{\rho^2+6a^2}\sqrt{\rho^2+9a^2}}\Bigg]\eta_1
+ e^{-i\psi/2}\alpha_3(\rho,\theta_1)\,\eta_3 + e^{-i\psi/2}\alpha_4(\rho,\theta_2)\,\eta_4 
\end{equation} 
with the functional forms of $\alpha_1, \alpha_3$ and $\alpha_4$ derived in Appendix B, see \eqref{alphas}. For large $\rho$ or
large $\lambda$, the behaviour of ${\hat G}_3^{(2,1)}$ is of the form:
\begin{equation}\label{large}
{\hat G}_3^{(2, 1)} ~ \to ~ {\rm constant} ~ + ~ {\rm log}~ \lambda
\end{equation}
and therefore ${\hat G}_3^{(2,1)}$ diverges logarithmically. This divergence is not problematic because eventually we are compactifying our 
manifold to a non-CY threefold. One should also observe that the ($2,1$) part of the fluxes in the original PT solution \cite{pt} asymptotically
goes to a constant. 

On the other hand, the asymptotic behaviour of the ($1,2$) part of the fluxes is more interesting. The explicit 
form of the ($1, 2$) part is given by:
\begin{equation}\label{12part}
{\hat G}_3^{(1, 2)} ~ = ~ \Bigg[\alpha_8 -\frac{27 P a^2}{\rho^3\sqrt{\rho^2+6a^2}\sqrt{\rho^2+9a^2}}\Bigg] \eta_8\,,
\end{equation}
with $\alpha_8$ given in \eqref{alphas}. Asymptotically ${\hat G}_3^{(1, 2)}$ now behaves in the following way:
\begin{equation}\label{small}
{\hat G}_3^{(1, 2)} ~ \to ~ \frac{1}{\lambda^2}
\end{equation}
and therefore goes to zero very fast. In fact the ($1, 2$) part of the fluxes in \cite{pt} also has the same behaviour 
asymptotically. 

Such an asymptotic behaviour of ${\hat G}_3^{(1, 2)}$ is good for us. 
This means that, since the fluxes vanish at the boundary, they should naturally exist once we compactify the resolved conifold to a compact threefold.
Furthermore we see that the ($1, 2$) part of the three form flux can be expressed alternatively as:
\begin{equation}\label{12form}
{\hat G}_3^{(1, 2)} ~ = ~ J \wedge \bar m
\end{equation}
with $\bar m$ being a ($0, 1$) form as one would have indeed expected. From our above consideration the ($0, 1$) 
form and $J$ are given in terms of the three one-forms in the following way:
\begin{equation}\label{jwbar}
\bar m ~ \equiv ~ h_1(\rho) \bar E_1 ~ = ~ 
\Bigg[\alpha_8 -\frac{27 P a^2}{\rho^3\sqrt{\rho^2+6a^2}\sqrt{\rho^2+9a^2}}\Bigg] \bar E_1, ~~~~~~~~~~
J ~ = ~ d \omega_1 ~ + ~ 4a^2 d \omega_\pm
\end{equation}
on the two patches $H_\pm$. The latter definition of $J$ is identical to the definition of $J$ in terms of the 
complex vielbeins $E_i$ given in \eqref{cs}\footnote{Note that the volume form is unique despite the existence of 
 multiple one-forms. The volume form is given by: $V = du \wedge dy \wedge d\zeta_+ = dv \wedge dz \wedge d\zeta_-$.}. 
It is also clear that the ($2, 1$) form cannot be expressed as 
\eqref{12form} using a one form because the ($2,1$) form is primitive. Observe, however, that 
the existence of a normalisable ($0, 1$) form doesn't always imply the existence of a non-trivial one-cycle in the 
manifold\footnote{Although, in the language of the fourfold the 
threefold base does have a non-vanishing first Chern class.}. 

Once we have the explicit ($1, 2$) forms, we still must see how this is uplifted in the M-theory picture. This is where 
things become somewhat subtle. The generic uplift of type IIB three-forms was given in \cite{DRS, GKP} in the following form:
\begin{equation}\label{gfourone}
G_4 ~ = ~ -\frac{{\hat G}_3 \wedge d\bar z}{\tau - \bar\tau} ~ 
+ ~ \frac{\bar {\hat G}_3 \wedge dz}{\tau - \bar\tau}
~ = ~ {\hat F}_3 \wedge dx^3 ~ + ~ {\hat H}_3 \wedge dx^{11}
\end{equation}
where we have used the usual definitions of $G_3$ and $dz$, namely: 
${\hat G}_3 = {\hat F}_3 - \tau {\hat H}_3$ and 
$dz = dx^{11} + \tau dx^3$ (although $d\tau \ne 0$). Thus $\widetilde{F}_3 = {\hat F}_3 - C_0 {\hat H}_3$ and 
${\hat F}_3 = d\hat C_2$ to comply with the notation used in section \ref{iib}.
With these definitions, the T-duality from IIB to M-theory
works in an expected way.

However, because of the presence of $d\bar z$ and $dz$ in \eqref{gfourone}, the uplift of a ($2, 1$) form is indeed 
a ($2, 2$) form, but the naive uplift of a ($1, 2$) form becomes a (1, 3) or a (3, 1) form, none of which are 
suitable for our case because these forms are ASD in M-theory. In the literature such subtlety was never observed 
because the ISD fluxes were never taken to have (1, 2) components. For our case, as we saw above, such forms are allowed because of their localised 
and normalisable nature. 

Indeed, such localisation of fluxes will eventually help us to show that the (1, 2) forms would also lift to F-theory 
as (2, 2) forms. To see this, observe that F-theory allows the following two important topological couplings:
\begin{equation}\label{topcop}
{\cal L}_1 ~ \equiv ~ \int_{{\cal M}_{12}} C_4 \wedge G_4 \wedge G_4, ~~~~~~~ 
{\cal L}_2 ~ \equiv ~ \int_{{\cal M}_{12}} G_4 \wedge G_4 \wedge G_4  \, ,
\end{equation}
where $C_4$ is the self-dual four-form in type IIB theory and ${\cal M}_{12}$ is the twelve dimensional space
(see \cite{feramin} and references therein for more details on these couplings).

The coupling ${\cal L}_1$ is well known. 
This leads to the standard Chern-Simmons term on D7 branes when we decompose the 
four-form as $G_4 = F \wedge d\omega$, where $d\omega$ is the normalisable two-form derived earlier and $F$ is the 
gauge flux on a D7 brane. The second coupling, ${\cal L}_2$, concerns us here. In type IIB there are no fundamental 
massless four-forms other than $C_4$ discussed above. How do we interpret $G_4$? The coupling that we 
are concerned with here is 
\begin{equation}\label{newcoupla}
\int_{{\cal M}_8} G_4 \wedge F \wedge F  \, ,
\end{equation}
where ${\cal M}_8$ is an eight dimensional surface. The only eight dimensional surface that we have in type IIB 
is the surface of the D7 brane. Therefore, we should expect the coupling \eqref{newcoupla} to show up 
on the surface of the D7 brane as some kind of {\it compact} four-form coupling to it. 

Existence of such compact four-forms can arise from the Chern-Simons terms on the D7 branes. One can easily see
that there is a coupling of the form:
\begin{equation}\label{cscoup}
\int_{{\cal M}_8} \widetilde{F}_3 \wedge {\cal A} \wedge F \wedge F \equiv 
\int_{{\cal M}_8} \left({\hat F}_3 - C_0 ~{\hat H}_3\right) \wedge {\cal A} \wedge F \wedge F
\end{equation}
when we choose the orientation of the D7 branes such that the arbitrary phase factor $\tilde\theta$ in 
\eqref{corembed} is a constant and our gauge invariant field on any D7 brane is ${\cal F} =\check B - F$ where 
$\check B$ is the pullback of the NS 2--form\footnote{We take $2\pi \alpha' = 1$ henceforth.}.

The above form of the coupling \eqref{cscoup} is of the type \eqref{newcoupla} provided the one-form ${\cal A}$ 
also becomes localised. Observe that both the three-forms appearing in \eqref{cscoup} are the localised (1,2) 
forms. Let us then assume that the one-form is ${\cal A } = l_1(\theta_1) dx^3$, 
where $l_1(\theta_1)$ is some localised
function that we will specify soon. We have also made a gauge choice to orient ${\cal A}$ along $x^3$ direction. With 
this we see that one choice of localised four-form flux is:
\begin{equation}\label{locfor} 
G_4^{(1)} ~ \equiv ~ l_1(\theta_1) \widetilde{F}_3 \wedge dx^3 ~ 
= ~ l_1(\theta_1) \left({\hat F}_3 \wedge dx^3 - C_0 ~{\hat H}_3 \wedge dx^3\right) \, .
\end{equation}
There is another choice of localised four-form flux that we can have in addition to \eqref{locfor}. This choice
can be motivated from the Born-Infeld terms of the D7 branes, and is given by:
\begin{equation}\label{locfortwo}
G_4^{(2)} ~ = ~ {\hat H}_3 \wedge \omega \, ,
\end{equation} 
where $\omega$ is the one-form derived in \eqref{oneform}. Once we compactify the internal space, the total 
axionic charge has to vanish. In that case both $G_4^{(1)}$ and $G_4^{(2)}$ simplify. In the presence of axion 
field, the total localised four-form flux is given by:
\begin{equation}\label{gfour}
G_4 ~ \equiv ~ G_4^{(1)} + G_4^{(2)} ~ = ~ {\hat H}_3 \wedge \omega +  
l_1(\theta_1) \left(-{\hat F}_3 \wedge dx^3 + 
C_0 ~{\hat H}_3 \wedge dx^3\right) \, ,
\end{equation}
which can be put in a very suggestive form if we define $l_1(\theta_1) = l(\theta_1)$ 
with $l(\theta_1)$ being the  
function of $\theta_1$ given in \eqref{ltheta} and \eqref{oneform}:
\begin{equation}\label{gfupp}
G_4~ = ~ l(\theta_1)\Big({\hat H}_3 \wedge dx^{11} -{\hat F}_3 \wedge dx^3 + 2C_0~{\hat H}_3 \wedge dx^3\Big)~ = ~ 
- l(\theta_1)~ \frac{{\hat G}^{(1, 2)}_3 \wedge dz}{\tau - \bar\tau} ~ + ~ {\rm c.c}
\end{equation}
with $dz = dx^{11} + \tau dx^3$ and ${\hat G}^{(1, 2)}_3$ being the (1, 2) form. 
The above four-form is clearly a (2, 2) form
as one would have expected from the earlier discussions \cite{beckersusy, haack, Dinerohm}. Notice however that the 
four-form flux is {\it not} closed. 

It is also interesting to note that since ${\hat G}_3^{(1, 2)}$ 
is of the form $J \wedge \bar m$ (see \eqref{12form}), the
localised (2, 2) form in M-theory becomes:
\begin{equation}\label{locg}
G_4~ \equiv ~ \frac{1}{2}~{\rm Re}~\Big(i e^\phi ~l(\theta_1) ~J \wedge \bar m \wedge dz\Big)
\end{equation}
At this point we may want to connect the four-form with the results given in \cite{beckersusy, haack}. The 
four-form should be related to ${\cal J} \wedge {\cal J}$ in M-theory where ${\cal J}$ is the fundamental
(1, 1) form for the fourfold. Defining ${\cal J} = J + dz \wedge d\bar z$, we have
\begin{equation}\label{jj}
{\cal J} \wedge {\cal J} ~ = ~ J \wedge J ~ + ~ 2 J \wedge dz \wedge d\bar z \, .
\end{equation}
It is easy to follow these fluxes to see how they appear in type IIB side. The second component in 
\eqref{jj} i.e $J \wedge dz \wedge d\bar z$ becomes a three-form field strength in T-dual type IIA theory: 
\begin{equation}\label{curF}
(\tau - \bar\tau)~J \wedge dx^3
\end{equation}
whose origin will be discussed in the next section. 
Similarly, the first component 
in \eqref{jj} ($J \wedge J$) becomes a five-form in type IIB side which has one component 
along $x^3$ direction and other components inside the threefold. This takes the form:
\begin{eqnarray}\label{five}\nonumber 
G_5 & = & \frac{\rho^3}{9} ~{\rm sin}~\theta_1~d\rho \wedge e_\psi \wedge d\phi_1 \wedge d\theta_1 \wedge dx^3 +  
\frac{\rho(\rho^2 + 6a^2)}{9}~{\rm sin}~\theta_2~d\rho \wedge e_\psi \wedge d\phi_2 \wedge d\theta_2 \wedge dx^3 \\
&& ~ + ~ \frac{\rho^2(\rho^2 + 6a^2)}{18} ~{\rm sin}~\theta_1~{\rm sin}~\theta_2~d\phi_1 \wedge d\theta_1
 \wedge d\phi_2 \wedge d\theta_2 \wedge dx^3 \, .
\end{eqnarray}   
This five-form (or the equivalent four-form) is 
strongly reminiscent of the four-form that we called $G_4^{(1)}$ in \eqref{locfor}, which does have one component
along $x^3$ direction. Indeed, the five-form\footnote{This is clearly non-vanishing because the underlying four-form
is not closed as we saw above.}:
\begin{equation}\label{fnow}
\frac{dl}{d\theta_1}~d\theta_1 \wedge {\hat F}_3 \wedge dx^3 ~ 
+ ~ \frac{l}{2} ~ \left(d{\hat G}_3 + d{\bar{\hat G}_3}\right)
\wedge dx^3
\end{equation} 
that we get from our background flux does match with \eqref{five}, but \eqref{fnow} has more terms than \eqref{five}.
This difference appears because, once we compactify our manifold, the fundamental form $J$ would change which, would
change the five-form \eqref{five}. 

The connection we have established here gives a stronger justification for why the cosmological constant should vanish in the bulk. It 
may be interesting to see if the arguments of \cite{Dinerohm} could be applied to our scenario also. This will be 
discussed elsewhere. 


\vskip.1in

\section{Applications to Cosmology}\label{cosmo}
\setcounter{equation}{0}

\subsection{Compactification and non-K\"ahlerity}

There remain issues that were given only partial attention in our earlier sections. The first such issue is the nature of a possible compactification of our background, which will certainly not be a Calabi--Yau, nor even K\"ahler. In the F-theory section we discussed that the six--dimensional base 
cannot be a Calabi--Yau manifold as it has a non-vanishing first Chern class. By reducing to IIA we can argue that the T--dual IIB background will indeed be non--K\"ahler. This construction follows the ones laid out in \cite{andrei, gtpaper1}.

The three form flux \eqref{curF}
that we get in type IIA will dissolve in the metric once we T-dualise to type IIB theory, making the background 
non-K\"ahler\footnote{In M-theory once $d{\cal J} \ne 0$ the four-form flux ${\cal J} \wedge {\cal J}$ is not 
closed. This is of course consistent with our choice of four-form flux \eqref{gfour}.}. Once the background is non-K\"ahler there would be {\it extra} sources of fluxes in addition to the fluxes that 
we mentioned in \eqref{gfour}, namely ``geometric flux''. 
One can replace the type IIB three form NS flux by 
\begin{equation}\label{comthree}
{\tilde H}_3 ~ \equiv~ {\hat H}_3 + i d(e^{-\phi}J) \, .
\end{equation}
This complexification of the three form flux is not new and has been observed earlier in heterotic compactifications 
\cite{het1, het2, het3, het4}, which in turn gave rise to a new superpotential in the heterotic theory \cite{het5, 
curiolust1}. An interesting observation here is that the type IIB background itself becomes non-K\"ahler now as 
compared to the heterotic background where the type IIB background was conformally K\"ahler. 

We also remarked on possible generalisations of the IIB superpotential in section \ref{cc}. It seems clear that the GVW superpotential will get corrected if the moduli space is enlarged by non--trivial one--forms. For the case of a background that is mirror to a Calabi--Yau with NS flux (so it acquires a non--trivial $T^3$ fibration when the mirror symmetry is interpreted as three T--dualities --- the NS B--field becomes part of the metric in the mirror manifold \cite{andrei}), a superpotential has been proposed \cite{berglund}. Whether or not this is suggestive for our case requires further study. Thus far, we have no reason to believe that our IIB manifold (globally) admits an SU(3) structure. The space of generalised Calabi--Yau manifolds is much larger, though some work on superpotentials in this case appeared in \cite{louisgrana1, iman, louisgrana, luca}. If we could infer that our IIB background admits an SU(3) structure, then it would be guaranteed to be complex \cite{andrew, grana, dallagata} if it preserved supersymmetry. However, in the presence of susy--breaking flux we cannot infer the structure of the manifold. A complex manifold would have the advantage to give us control over the complex structure deformations.

\subsection{Inflationary dynamics} 

The major motivation for constructing the background in 
this paper was to study a model of inflation that may give slow roll dynamics with less fine tunings than the usual $D3-\overline{D3}$ scenarios \cite{KKLMMT, bdkm}. Let us therefore sketch a 
possible model of inflation using the resolved conifold background with D7 branes and additional D3 branes. 

Recall that D3/D7 inflation has primarily been studied in toroidal manifolds (see \cite{DHHK} and citations therein) of the 
form $T^n/{\Gamma}$ of which $K3 \times T^2/\mathbb{Z}_2$ is a 
special case. The F-term and D-term potentials appearing from the 
gaugino condensate and susy breaking fluxes, respectively, conspired to 
give a consistent resolution of the anomalies associated with the 
FI terms. 

We outline a possible scenario to achieve 
slow roll inflation when we combine the ideas of $D3-\overline{D3}$ in the ``warped throat'' (KKLMMT \cite{KKLMMT}) with D3/D7 models  \cite{DHHK})\footnote{Similar idea has been proposed independently by Cliff Burgess.}. 
We want to balance a D3 that is attracted towards the D7 (because of the non-primitive flux on the D7 worldvolume) with another force that drives the D3 toward the tip. This can be achieved by placing an anti-D3 there or by using a background in which the addition of a D3 explicitly breaks supersymmetry, such as the resolved warped deformed conifold \cite{DKS}. The motion of D3--branes towards the tip in the latter background is a consequence of the 
running dilaton. However, this potential alone is still too steep for slow--roll inflation. 

Combining both forces, however, we might hope to slow down the motion of the D3 in either the one or the other direction. There are two possible scenarios, depending on which force dominates:
\begin{itemize}
\item The D--term potential created by the non--primitive flux dominates and attracts the D3--brane towards the 
  wrapped D7 brane. Inflation ends when the D3 dissolves into the D7 as non-commutative instantons and supersymmetry is restored.
\item The attraction towards the anti--D3 brane at the bottom of the throat (or possibly a running dilaton in a more general background)   dominates. Inflation ends as all or some D3 branes getting annihilated by the 
anti--D3 brane(s) at the tip of the throat.
\end{itemize}
Naively one might hope that the motion would 
be slow because the D3 branes
are pulled in both directions. However, it may also turn out that the initial position of the D3 has to be heavily fine--tuned in this setup.

The F-term potential associated with the motion of the D3--branes towards the 
tip of the throat has recently been computed with the inclusion of holomorphically embedded D7--branes \cite{bdkm, axel, BCDF} using the 
analysis of \cite{BDKMMM}. 
If we want to combine the D-term and F-term potentials we are faced with an issue pointed out by \cite{nilles}: for a supersymmetric background it is impossible to have a $D$-term potential 
that could be used to pull the D3 brane towards the D7 branes. Thus if we want to switch on non--primitive
fluxes on the wrapped D7 branes we have to embed the D7 branes in a non-supersymmetric 
background. Our 
problem becomes threefold:
\begin{itemize}
\item Construct a supergravity background with embedded D7 branes that breaks supersymmetry 
  spontaneously.
\item Allow for a possible D-term uplifting by avoiding the no-go theorem of 
  \cite{nilles}, as pointed out by \cite{fernando}. Note that the D7 worldvolume theory will not only contribute the D but also possible F-terms, such that the issue of \cite{nilles} might be resolved.
\item Balance the D3 brane using the two forces: one from the D-term potential and the other 
  from the attractive force at the tip of the deformed conifold in the KKLMMT setup.
\end{itemize}
In this paper we have addressed the first two problems by constructing a non-supersymmetric background with 
D-terms on the D7 branes given by the pullback of a non--primitive flux. To analyse the last problem, we might have to go to a more generic 
background of the form given earlier as \eqref{resdefmetric} which 
is a resolved warped deformed conifold with $F_1 \ne F_2$ and $b\ne 0$, i.e. both the two and the 
three cycles are non vanishing.  
Most of the literature deals with the limit where the manifold looks like a singular conifold.
This isn't the most generic situation so we have to 
go away from the usual conifold background. However, taking a resolved warped deformed conifold creates non-trivial 
dilaton profile from two sources now:
\begin{itemize}
\item From the D7 branes, and
\item  From the unequal sizes of the two-cycles.
\end{itemize}
The running of dilaton from the first case can already be seen at a supersymmetric level in the Ouyang background
\cite{ouyang}, which was originally analysed for a non--compact singular conifold background. Once we blow up 
resolution cycles of the conifold and switch on fluxes, the second case mentioned above kicks in, and 
we must discuss the combined effects to get the full background geometry. This makes the problem much harder to 
solve and will be tackled in the sequel to this paper. 

The warped resolved conifold however may still be a good model of inflation with D-term uplifting. We would have to extend our analysis beyond the case $\mu=0$ (in this case the D7 extend all the way down the throat, which would not allow us to place a D3 \emph{between} the D7 and the tip) and to other embeddings, such as the Kuperstein embedding \cite{kuperstein}. Our preliminary analysis indicates that the value of the D-terms should depend on the choice of embedding.

\subsection{Supersymmetry restoration}

When the D3--brane falls into the D7--branes at the end of inflation
we expect supersymmetry to be restored. Such a susy restoration was first described 
in \cite{DHHK}. For our case, the situation is more involved. From the F-theory point of view, the {\it total} G-flux at the 
end of the inflation can be succinctly presented as: 
\begin{equation}\label{totg}
G_{\rm total} ~\equiv ~ G^{(2,2)}_{P} ~ + ~ c_1 ~{\cal J} \wedge {\cal J} ~ + ~ c_2 ~ F\wedge d\omega ~ + ~ 
c_3~ {\hat H}_3 \wedge \omega \, ,
\end{equation}
where $c_i$ are some defined functions of the coordinates ($\theta_1, \phi_1$) or ($\theta_2, \phi_2$) depending on 
which branch \eqref{sevencoor} we are on, 
$G^{(2,2)}_{P}$ is the primitive
part of the $G$-flux that come from the uplift of the type IIB (2, 1) forms,   
and $F$ is the gauge flux 
induced by dissolving the D3 brane inside the D7 branes. 
The 1-form $\omega$ was defined in \eqref{oneform}. The last term coming from the ${\hat H}_3$ coupling is 
non-primitive, and because of that
in the absence of $F$ flux, the $G$ flux was 
(2, 2) but non-primitive. Observe that in the presence of $F$ flux we can in fact demand:
\begin{equation}\label{primgfll}
{\cal J} \wedge G_{\rm total} ~ = ~ 0
\end{equation}
and therefore restore supersymmetry with (2,2) fluxes.

The $F$ flux used to restore supersymmetry in the above paragraph could be interpreted in two ways: switching 
on second Chern class or switching on first Chern class. The former, which leads to instantons, is the end point of the 
D3 brane dissolving on the D7 branes. 
The latter, however, gives rise to a bound state of a D5 brane with the D7 branes. 
Such a technique of restoring supersymmetry has already been discussed in \cite{Dgwyn1, Dgwyn2} and could 
probably be used to restore supersymmetry in the limit where the resolution parameter $a$ goes to zero. This 
would then be one simple way of restoring supersymmetry in the original Ouyang construction \cite{ouyang}. 

\section{Conclusions and future directions}
\setcounter{equation}{0}

Related to our flux choice is another issue that deserves mentioning. The (1, 2) flux that we choose is 
ISD and solves equations of motion. One may also choose AISD flux if one changes the ans\"atze for the 
background geometry, i.e. if one ventures beyond conformal Calabi--Yau compactifications. Typically one can show that a compact conformally Calabi-Yau background only allows 
ISD fluxes that are also primitive. As we saw above, non-primitive ISD fluxes are allowed on a compact 
non-K\"ahler background or on a non-compact Calabi-Yau background. 
However AISD fluxes are generically part of the solution to the equations of motion on non-K\"ahler backgrounds. Some recent papers 
dealing with this are \cite{uranga1, uranga2, kachrunew}.

One other question would be to reconcile the following puzzle\footnote{We thank Shamit Kachru for discussions 
on this point.}: 
A non-compact Calabi-Yau background could be 
dual to a gauge theory via gauge/gravity duality. In the gravity side it is possible to have supersymmetry breaking
without generating a cosmological constant. However on the gauge theory side it is in general {\it not} possible to 
break supersymmetry keeping the cosmological constant zero. 
Maybe in our case there is some obstruction to finding a gauge theory that is dual to our non--compact background.
In fact the non-compact resolved conifold background that we took is dual (in the sense of a geometric transition) to a pure supergravity background\footnote{A warped deformed conifold with fluxes.} 
if we consider  wrapped D5 branes instead of the RR three form fluxes\cite{vafa1, civ, gtpaper1, gtpaper2}. The resolved conifold as a supergravity background without branes is only known to be dual 
to a gauge theory in IIA when there are one form gauge fluxes present. On the other hand, once susy is restored via 
one of the possible mechanism discussed earlier, our background could in principle be dual to some gauge theory.
The other known duality is the one studied 
recently in \cite{klebumuru} that uses large number of D3 branes in the resolved conifold background. This model
is very different from the one studied by us here. 

In summary, we have applied the methods of \cite{ouyang} to the warped resolved conifold background of Pando--Zayas and Tseytlin \cite{pt}. We found a supergravity background that breaks supersymmetry spontaneously due to fluxes of type (1,2) without generating a bulk cosmological constant. The pullback of the NS B--field onto the D7--worldvolume gives rise to D-terms, which vanish in the limit of vanishing resolution parameter $a\to 0$, i.e. when we approach the original singular background of \cite{ouyang}. We have also convinced ourselves that the worldvolume flux in the original embedding is non-primitive and should therefore break supersymmetry. A cancellation of this effect by adding gauge fluxes would be worth further study. We should then also re-examine the D-terms we find on the resolved conifold. In the case we studied, the D7 gauge fluxes were zero and the D-terms entirely due to the non--primitive NS B--field. In general we would also expect F--terms from the D7 worldvolume theory.
As soon as we want to apply our model to inflationary model building, we would want to add D3--branes into the picture. This gives rise to new degrees of freedom and further influences the F-terms.

In parallel to the IIB discussion we have also studied the F-theory lift of our background. We showed how the non--primitive ISD $G_3$--flux lifts to non--primitive selfdual $G_4$ flux, 
which should be proportional to $J\wedge J$. We gave an explicit construction of the normalisable harmonic forms that correspond to the D7--branes. These harmonic forms would appear as second cohomologies of the compactified fourfold. 
We showed how a compact non-K\"ahler threefold base could be constructed which would have the required 
local features of a resolved conifold background that we studied for the type IIB scenario. A more detailed account of the
fourfold, including its cohomological structure, will be discussed in the future.

\vspace*{0.5cm}
\subsection*{Acknowledgements}

We are most grateful to Cliff Burgess, Jim Cline, Hassan Firouzjahi,
Andrew Frey, Hans Jockers, Shamit Kachru, Peter Ouyang, Gary Shiu, Radu Tatar and Angel Uranga for helpful discussions.
A.K. would like to thank the KITPC in Beijing for hospitality during part of this work and the organisers of the ``String Theory and Cosmology'' workshop for creating an enjoyable and stimulating atmosphere. P.F. would like to thank the TIFR in Mumbai and the organisers of the ``From Strings to LHC - II" advanced school and conference, for hospitality during part of this work. The work of all authors is supported by the Natural Sciences and Engineering Research Council of Canada (NSERC).
\vspace*{0.5cm}

\newpage
\begin{appendix}
\def\theequation{\Alph{section}.\arabic{equation}}

\section{The geometry of the resolved conifold}\label{rescone}

\setcounter{equation}{0}

The resolved conifold is a manifold which looks asymptotically like the singular conifold, but is non--singular at the tip. Its geometry can be derived by starting with the singular version, a non--compact Calabi--Yau 3--fold, that can be embedded in $\mathbb{C}^4$ as \cite{candelas}
\begin{equation}
  \sum_{i=1}^4 z_i^2 \,=\, 0\,.
\end{equation}
This describes a cone over $S^2\times S^3$, which becomes singular at the origin.
By a change of coordinates this can also be written as
\begin{equation}
  yz-uv \,=\, 0\,,
\end{equation}
which is equivalent to non--trivial solutions to the equation 
\begin{equation}
\begin{pmatrix} z & u \\ v & y\end{pmatrix}
\begin{pmatrix} \xi_1 \\ \xi_2 \end{pmatrix}
\,=\, 0\,,
\end{equation}
in which $\xi_1, \xi_2$ are homogeneous coordinates on $\mathbb{CP}^1\sim S^2$. For $(u,v,y,z)\ne 0$ (away from the tip), they describe again a conifold. But at $(u,v,w,z)=0$ this is solved by any pair $(\xi_1,\xi_2)$. Due to the overall scaling freedom $(\xi_1,\xi_2)\sim (\lambda\xi_1,\lambda\xi_2)$ we can mod out by this equivalence class and $(\xi_1,\xi_2)$ actually describe a $\mathbb{CP}^1\sim S^2$ at the tip of the cone. 
The resolved conifold can be covered by two complex coordinate patches ($H_+$ and $H_-$), given by 
\begin{eqnarray}\label{coordpatch}
H_+ &=& \{ \xi_1 \neq 0 \} = \{ (u,y;\lambda) | u,y,\lambda \in \mathbb{C} \} \ , \ \lambda = \frac{\xi_2}{\xi_1}   \\
H_- &=& \{ \xi_2 \neq 0 \} = \{ (v,z;\mu) | v,z,\mu \in \mathbb{C} \} \ , \ \mu = \frac{\xi_1}{\xi_2} \, . 
\end{eqnarray}
On $H_+$ we have that 
\begin{equation}
z = -u \lambda \ , v = - y \lambda  \,,
\end{equation}
on $H_-$
\begin{equation}
y = -v \mu \ , u = - z \mu  \,,
\end{equation}
and on the intersection of these two patches, the coordinates are related by
\begin{equation}\label{overlap}\nonumber
(v,z;\mu) = (-y\lambda,-u\lambda ; 1/\lambda)  \,.
\end{equation}
The holomorphic coordinates are conveniently parameterised by
\begin{eqnarray}\label{holocoord}\nonumber
  z & =&  \left ( 9 a^2 \rho^4 + \rho ^6 \right ) ^{1/4} e^{i/2(\psi-\phi_1-\phi_2)}\,\sin(\theta_1/2)\sin(\theta_2/2) \\ \nonumber
  y & =&  \left ( 9 a^2 \rho^4 + \rho ^6 \right ) ^{1/4} e^{i/2(\psi+\phi_1+\phi_2)}\,\cos(\theta_1/2)\cos(\theta_2/2) \\
  u & =&  \left ( 9 a^2 \rho^4 + \rho ^6 \right ) ^{1/4} e^{i/2(\psi+\phi_1-\phi_2)}\,\cos(\theta_1/2)\sin(\theta_2/2) \\ \nonumber
  v & =&  \left ( 9 a^2 \rho^4 + \rho ^6 \right ) ^{1/4} e^{i/2(\psi-\phi_1+\phi_2)}\,\sin(\theta_1/2)\cos(\theta_2/2)\,.
\end{eqnarray}
Here, $\theta_i=0\ldots \pi$, $\phi_i=0\ldots 2\pi$ are the usual Euler angles on $S^2$, $\psi=0\ldots 4\pi$ describes a U(1) fibre over the two 2--spheres and $\rho=0\ldots\infty$ is the radial coordinate. Note that our radial coordinate $\rho$ is related to the commonly used $r$ via $\rho^2=3/(2 r^2) F'(r^2)$, where $F(r^2)$ appears in the K\"ahler potential $K$ of the resolved conifold
\begin{equation}
  K(r) \,=\, F(r^2)+4a^2\log(1+|\lambda|^2)\,.
\end{equation}
Note that the K\"ahler potential is not a globally defined quantity, since $\lambda$ is only defined on the patch $H_+$ that excludes $\xi_1=0$. For completeness let us also quote \cite{candelas, pt}
\begin{eqnarray}\label{Fprime}
  F'(r^2) \,=\, \frac{\partial F(r^2)}{\partial r^2} &=& \frac{1}{r^2}\,\left(-2a^2 + 4a^2 N^{-1/3}(r) + N^{1/3}(r)\right) \quad\mbox{with}\\ \label{defN}
  N(r) &=& \frac{1}{2}\,\left(r^4-16 a^2+\sqrt{r^8-32a^6r^4}\right)\,.
\end{eqnarray}
The inverse relation between $\rho$ and $r$ is found to be
\begin{equation}\label{rhoandr}
  r \,=\, \left(\frac{2}{3}\right)^{3/4}\,(9a^2\rho^4+\rho^6)^{1/4}\,.
\end{equation}

\noindent In terms of these real coordinates the Ricci--flat K\"ahler metric on the resolved conifold reads
\begin{eqnarray}\label{resmetric}\nonumber
  ds^2_{\rm res} & = & \kappa(\rho)^{-1}\,d\rho^2 + \frac{\kappa(\rho)}{9}\,\rho^2\big(d\psi+\cos\theta_1\,d\phi_1
    +\cos\theta_2\,d\phi_2\big)^2 \\
  & &+ \frac{\rho^2}{6}\,\big(d\theta_1^2+\sin^2\theta_1\,d\phi_1^2\big) 
    +\frac{\rho^2+6a^2}{6}\,\big(d\theta_2^2+\sin^2\theta_2\,d\phi_2^2\big)\,,
\end{eqnarray}
with $\kappa(\rho)=(\rho^2+9a^2)/(\rho^2+6a^2)$.  In the limit $a\to 0$ one recovers the singular conifold metric, therefore $a$ is called ``resolution'' parameter and gives the radius of the blown--up 2--sphere at the tip. 

\noindent It will be useful later on to have a set of vielbeins that describes this metric, i.e.
\begin{equation}
  ds^2 \,=\, \sum_{i=1}^6 (e_i)^2\,.
\end{equation}
Following \cite{ankenon} we choose
\begin{eqnarray}\label{resvielb}\nonumber
  e_1 &=& \kappa^{-1/2}\,d\rho\\ \nonumber
  e_2 &=& \frac{\rho\sqrt{\kappa}}{3}\,(d\psi+\cos\theta_1\,d\phi_1+\cos\theta_2\,d\phi_2) 
    \,=\, \frac{\rho\sqrt{\kappa}}{3}\,e_\psi\\ \nonumber
  e_3 &=&\frac{\rho}{\sqrt{6}}\,(\sin\psi/2\,\sin\theta_1\,d\phi_1+\cos\psi/2\,d\theta_1)\\ 
  e_4 &=&\frac{\rho}{\sqrt{6}}\,(-\cos\psi/2\,\sin\theta_1\,d\phi_1+\sin\psi/2\,d\theta_1)\\ \nonumber
  e_5 &=&\frac{\sqrt{\rho^2+6a^2}}{\sqrt{6}}\,(\sin\psi/2\,\sin\theta_2\,d\phi_2+\cos\psi/2\,d\theta_2)\\ \nonumber
  e_6 &=&\frac{\sqrt{\rho^2+6a^2}}{\sqrt{6}}\,(-\cos\psi/2\,\sin\theta_2\,d\phi_2+\sin\psi/2\,d\theta_2)
\end{eqnarray}
as they lead to a closed K\"ahler form $J$ as well as a closed holomorphic 3--form $\Omega$ with a simple complex structure induced by
\begin{equation}\label{J}
  J^{(1,1)} \,=\, e_1\wedge e_2+ e_3\wedge e_4+e_5\wedge e_6\,,\qquad
  \Omega^{(3,0)} \,=\, (e_1+ie_2)\wedge(e_3+ie_4)\wedge(e_5+ie_6)\,,
\end{equation}
in other words we define our complex vielbeins to be
\begin{equation}\label{cs}
  E_1 \,=\, e_1+i\,e_2\,,\qquad  E_2 \,=\, e_3+i\,e_4\,,\qquad  E_3 \,=\, e_5+i\,e_6\,. 
\end{equation}
This results in a coordinate expression for $J$ as
\begin{eqnarray}\label{Jres}\nonumber
  J &=& \frac{\rho}{3}\,d\rho\wedge (d\psi+\cos\theta_1\,d\phi_1+\cos\theta_2\,d\phi_2)\\
  & & + \frac{\rho^2}{6}\,\sin\theta_1\,d\phi_1\wedge d\theta_1+ \frac{\rho^2+6a^2}{6}\,\sin\theta_2\,d\phi_2\wedge 
    d\theta_2\,.
\end{eqnarray}

\section{Ouyang embedding of D7--branes on the resolved conifold}\label{embed}
\setcounter{equation}{0}

In this appendix we describe how D7--branes can be embedded in the PT background. We use the Ouyang \cite{ouyang} embedding 
\begin{equation}
  z \,=\, \mu^2\,,
\end{equation}
where $z$ is one of the holomorphic coordinates defined in \eqref{holocoord}. While this choice was orginally made for the singular conifold, it continues to give a consistent holomorphic embedding on both patches. From \eqref{overlap}, it is clear that selecting $z=\mu^2$ on $H_-$ implies that $-u \lambda=\mu^2$ on the intersection with $H_+$, which consistently gives $z=\mu^2$ on all of $H_+$. 

While the case $\mu \neq 0$, where the D7-brane does not extend to the tip of the throat, is of primary interest for inflationary models, we set $\mu=0$ for simplicity of calculation. As a consistency check we should always be able to recover a supersymmetric solution in the limit $a\to 0$.
The D7--brane induces a non--trivial axion--dilaton
\begin{equation}\label{dilbehav}
  \tau \,=\,  \frac{i}{g_s}+\frac{N}{2\pi i} \log z\,,
\end{equation} 
where N is the number of embedded D7-branes. 
Our goal is to determine the change the dilaton induces in the other fluxes and the warp factor. We will closely follow the method laid out in \cite{ouyang} and solve the SuGra equation of motion only in linear order $g_sN$. That said, we neglect any backreaction onto the geometry beyond a change in the warp factor, i.e. we will assume the manifold remains a conformal resolved conifold.

Consider first the Bianchi identity, which in leading order becomes ($H_3$ indicates the unmodified NS flux from \eqref{fluxres}, whereas the hat indicates the corrected flux at leading order)
\begin{eqnarray}
  d\hat G_3 & =& d\hat F_3 - d \tau \wedge \hat H_3 - \tau \wedge d\hat H_3 = -d \tau \wedge H_3 +\mathcal{O}((g_s N)^2)  \\ \nonumber
  & = & -\bigg( \frac{N}{2 \pi i } \frac{dz}{z} \bigg) \wedge \big( df_1(\rho) \wedge  d\theta_1 \wedge \sin\theta_1\,d\phi_1 
    + df_2(\rho) \wedge  d\theta_2 \wedge \sin\theta_2\,d\phi_2 \big) +\mathcal{O}((g_s N)^2)\,.
\end{eqnarray}
In order to find a 3--form flux that obeys this Bianchi identity, we make an ansatz
\begin{equation}
  \hat G_3 \,=\, \sum \alpha_i\,\eta_i
\end{equation}
where $\{\eta_i\}$ is a basis of imaginary self--dual (ISD) 3--forms on the resolved conifold given in \eqref{eta}.
We find a particular solution in terms of only four of above eight 3--forms
\begin{eqnarray}\label{particular1}
  P_3 &=& \alpha_1(\rho)\,\eta_1 + e^{-i\psi/2}\alpha_3(\rho,\theta_1)\,\eta_3 + e^{-i\psi/2}\alpha_4(\rho,\theta_2)\,\eta_4 
    +\alpha_8(\rho)\,\eta_8\,,
\end{eqnarray}
with 
\begin{eqnarray}\nonumber
  \alpha_3 &=& -3\sqrt{6} g_sNP\,\frac{72 a^4-3\rho^4+a^2\rho^2(\log(\rho^2+9 a^2)-56\log \rho)}
    {8\pi\rho^3(\rho^2+6 a^2)^2}\,\cot\frac{\theta_1}{2}\\[1ex]
  \alpha_4 &=& -9\sqrt{6} g_sNP\,\frac{\rho^2-9a^2\log(\rho^2+9 a^2)}{8\pi\rho^4\sqrt{\rho^2+6 a^2}}\,
    \cot\frac{\theta_2}{2}\\[1ex] \nonumber
  \alpha_8 &=& \frac{3a^2}{\rho^2+3a^2}\left[3g_sNP\frac{-9(\rho^2+4a^2)+28\rho^2\log\rho+(81a^2+13\rho^2)\log(\rho^2+9a^2)}
    {8\pi\rho^3\sqrt{\rho^2+6a^2}\sqrt{\rho^2+9a^2}}+\alpha_1(\rho)\right]
\end{eqnarray}
Note that $a_8$ is implicitly given by $\alpha_1$, which in turn is determined via the first order ODE
\begin{eqnarray}\nonumber
  \alpha_1'(\rho) &=& \frac{-3}{\rho(\rho^2+3a^2)(\rho^2+9a^2)\sqrt{\rho^2+6a^2}}\,\left[\frac{(162a^6+78a^4\rho^2+15a^2\rho^4+\rho^6)}
    {\sqrt{\rho^2+6a^2}}\,\alpha_1(\rho)\right.\\[1ex]
  & & \left.\; + 3g_sNP\frac{-162a^6+99a^4\rho^2+63a^2\rho^4+6\rho^6 +14a^2\rho^2(\rho^2+9a^2)\log\frac{\rho^2}{\rho^2+9a^2}}{4\pi\rho^3\sqrt{\rho^2+9a^2}}\right]\,.
\end{eqnarray}
Letting $a\to 0$ in above equations, we do indeed recover the singular conifold solution of \cite{ouyang}. Keeping the resolution parameter $a$ finite instead, we can solve for $\alpha_1(\rho)$ 
\begin{equation}
   \alpha_1(\rho) \,=\, \frac{3g_sNP}{8\pi\rho^3}\frac{ \left[18 a^2 - 36(\rho^2+3a^2)\log\left(\frac{\rho}{a}\right) 
    + (10\rho^2+72a^2)\log\left(\frac{\rho^2}{\rho^2+9a^2}\right)\right]} {\sqrt{\rho^2+6a^2}\sqrt{\rho^2+9a^2}}
\end{equation}
Furthermore, we find a homogeneous solution
\begin{eqnarray}
  G_3^{hom} &=& \beta_1(z,\rho)\,\eta_1 + e^{-i\psi/2}\beta_3(\rho,\theta_1)\,\eta_3 
    + e^{-i\psi/2}\beta_4(\rho,\theta_2)\,\eta_4 \\ \nonumber 
  & &  + e^{-i\psi}\beta_5(\rho,\theta_1,\theta_2)\,\eta_5 +\beta_8(z,\rho)\,\eta_8\,,
\end{eqnarray}
with
\begin{eqnarray}\label{betas}\nonumber 
  \beta_1 &=& \frac{-3i}{8\rho^3\sqrt{\rho^2+6a^2}\sqrt{\rho^2+9a^2}}\,\big[12(\rho^2+3a^2)\log z+18a^2+10(\rho^2-9a^2)\log \rho\\ \nonumber
    & & \phantom{8\rho^3\sqrt{\rho^2+6a^2}\sqrt{\rho^2+9a^2}} +(13\rho^2+99a^2)\log(\rho^2+9a^2)\big]\\[1ex] \nonumber
  \beta_3 &=& 3i\sqrt{6}\,\left(\frac{-36a^4+3\rho^4+2a^2\rho^2\big(20\log\rho-\log(\rho^2+9a^2)\big)}{4\rho^3(\rho^2+6a^2)^2}\right)\,
    \cot\frac{\theta_1}{2}\\[1ex]
  \beta_4 &=& -9i\sqrt{6}\,\left(\frac{\rho^2-6a^2\log(\rho^2+9a^2)}{4\rho^4\sqrt{\rho^2+6a^2}}\right)\,\cot\frac{\theta_2}{2}\\[1ex] \nonumber
  \beta_5 &=& \frac{-9i\,(\cot\frac{\theta_1}{2}\,\cos\theta_2+\cot\theta_1)}{2\rho^2\sqrt{\rho^2+9a^2}\sin\theta_2}\\[1ex]
    \nonumber
  \beta_8 &=& \frac{-27ia^2}{8\rho^3\sqrt{\rho^2+6a^2}\sqrt{\rho^2+9a^2}}\,\big[4\log z+6-10\log \rho-\log(\rho^2+9a^2)\big]
\end{eqnarray}
This solution has the right singularity structure at $z=0$ and $\rho=0$, but it does not transform correctly under $SL(2,\mathbb{Z})$; only the particular solution does. We therefore conclude that the correction to the 3--form flux, which is in general a linear combination of $P_3$ and $G_3^{hom}$, is given by \eqref{particular1} only
\begin{equation}
  \hat{G}_3 \,=\, G_3+P_3\,.
\end{equation}
We can now determine the change in the remaining fluxes and the warp factor, at least to linear order in $(g_sN)$. 
We find the corrected fluxes from the equations
\begin{equation}
  \hat H_3 \,=\, \frac{ \overline{\hat{G}}_3 - \hat{G}_3 }{\tau - \bar{\tau}}\qquad\mbox{and}\qquad 
    \widetilde{F}_3 \,=\, \frac{\hat{G}_3 + \overline{\hat{G}}_3}{2}\,,
\end{equation}
which evaluates to
\begin{eqnarray}\nonumber
  \hat H_3 &=& d\rho\wedge e_\psi\wedge(c_1\,d\theta_1+c_2\,d\theta_2) + d\rho\wedge(c_3\sin\theta_1\,d\theta_1\wedge d\phi_1-c_4\sin\theta_2\,d\theta_2\wedge d\phi_2)\\
  & & +\left(\frac{\rho^2+6a^2}{2\rho}\,c_1\sin\theta_1\,d\phi_1 
    -\frac{\rho}{2}\,c_2\sin\theta_2\,d\phi_2\right)\wedge d\theta_1\wedge d\theta_2\,,\\ \nonumber
  \widetilde{F}_3 &=& -\frac{1}{g_s}\,d\rho\wedge e_\psi\wedge(c_1\sin\theta_1\,d\phi_1+c_2\sin\theta_2\,d\phi_2)\\ \nonumber
  & &  +\frac{1}{g_s}\,e_\psi\wedge(c_5\sin\theta_1\,d\theta_1\wedge d\phi_1-c_6\sin\theta_2\,d\theta_2\wedge d\phi_2)\\ 
  & & -\frac{1}{g_s}\,\sin\theta_1\sin\theta_2\left(\frac{\rho}{2}\,c_2 \,d\theta_1-\frac{\rho^2+6a^2}{2\rho}\,c_1\,d\theta_2 \right)
    \wedge d\phi_1\wedge d\phi_2\,.
\end{eqnarray}
We have introduced the coefficients
\begin{eqnarray}\label{defc}\nonumber
  c_1 &=& \frac{g_s^2PN}{4\pi\rho(\rho^2+6a^2)^2}\,\big(72 a^4-3\rho^4-56a^2\rho^2\log\rho+a^2\rho^2\log(\rho^2+9a^2)\big)\,
    \cos\frac{\theta_1}{2}\\ 
  c_2 &=& \frac{3g_s^2PN}{4\pi\rho^3}\,\big(\rho^2-9a^2\log(\rho^2+9a^2)\big)\, \cos\frac{\theta_2}{2}\\ \nonumber
  c_3 &=& \frac{3g_sP\rho}{\rho^2+9a^2}+\frac{g_s^2PN}{8\pi\rho(\rho^2+9a^2)}\,\Big[-36a^2-36\rho^2\log a+34\rho^2\log\rho\\ \nonumber
    & & \qquad\qquad\qquad\qquad +(10\rho^2+81a^2)\log(\rho^2+9a^2)+12\rho^2\log\left(\sin\frac{\theta_1}{2}\sin\frac{\theta_2}{2}\right)\Big]\\  
    \nonumber
  c_4 &=& \frac{3g_sP(\rho^2+6a^2)}{\kappa\rho^3}+\frac{g_s^2NP}{8\pi\kappa\rho^3}\,\Big[18a^2-36(\rho^2+6a^2)\log a+(34\rho^2+36a^2) 
    \log\rho\\ \nonumber 
  & & \qquad\qquad + (10\rho^2+63a^2)\log(\rho^2+9a^2)+(12\rho^2+72a^2)\log\left(\sin\frac{\theta_1}{2}\sin\frac{\theta_2}{2}\right)\Big]
\end{eqnarray}
\begin{eqnarray}\nonumber
  c_5 &=& g_sP+\frac{g_s^2PN}{24\pi(\rho^2+6a^2)}\,\Big[18a^2-36(\rho^2+6a^2)\log a+8(2\rho^2-9a^2)\log\rho\\ \nonumber
  & & \qquad\qquad\qquad\qquad\qquad\qquad\qquad\qquad\qquad\qquad\;\;+(10\rho^2+63a^2)\log(\rho^2+9a^2)
    \Big]\\ \nonumber
  c_6 &=& g_sP+\frac{g_s^2PN}{24\pi\rho^2}\,\Big[-36a^2-36\rho^2\log a+16\rho^2\log\rho+(10\rho^2+81a^2)\log(\rho^2+9a^2)\Big]
\end{eqnarray}
This allows us to write the NS 2--form potential
\begin{eqnarray}
  B_2 &=& \left(b_1(\rho)\cot\frac{\theta_1}{2}\,d\theta_1+b_2(\rho)\cot\frac{\theta_2}{2}\,d\theta_2\right)\wedge e_\psi\\ \nonumber
  & + &\left[\frac{3g_s^2NP}{4\pi}\,\left(1+\log(\rho^2+9a^2)\right)\log\left(\sin\frac{\theta_1}{2}\sin\frac{\theta_2}{2}\right)
    +b_3(\rho)\right]\sin\theta_1\,d\theta_1\wedge d\phi_1\\ \nonumber 
  & - & \left[\frac{g_s^2NP}{12\pi\rho^2}\left(-36a^2+9\rho^2+16\rho^2\log\rho+\rho^2\log(\rho^2+9a^2)\right)
    \log\left(\sin\frac{\theta_1}{2}\sin\frac{\theta_2}{2}\right)+b_4(\rho)\right]\\ \nonumber
  & & \qquad\qquad \times \sin\theta_2\,d\theta_2\wedge d\phi_2
\end{eqnarray}
with the $\rho$-dependent functions 
\begin{eqnarray}\label{defb}\nonumber
  b_1(\rho) &=& \frac{g_S^2NP}{24\pi(\rho^2+6a^2)}\big(18a^2+(16\rho^2-72a^2)\log\rho+(\rho^2+9a^2)\log(\rho^2+9a^2)\big)\\
  b_2(\rho) &=& -\frac{3g_s^2NP}{8\pi\rho^2}\big(\rho^2+9a^2\big)\log(\rho^2+9a^2)
\end{eqnarray}
and $b_3(\rho)$ and $b_a(\rho)$ are given by the first order differential equations
\begin{eqnarray}\nonumber
  b_3'(\rho) &=& \frac{3g_sP\rho}{\rho^2+9a^2} + \frac{g_s^2NP}{8\pi\rho(\rho^2+9a^2)}\Big[-36a^2-36a^2\log a+34\rho^2\log\rho\\ \nonumber
  & & \qquad\qquad\qquad\qquad\qquad\qquad+(10\rho^2+81a^2)
    \log(\rho^2+9a^2)\Big]\\ 
  b_4'(\rho) &=& -\frac{3g_sP(\rho^2+6a^2)}{\kappa\rho^3} - \frac{g_s^2NP}{8\pi\kappa\rho^3}\Big[18a^2-36(\rho^2+6a^2)\log a\\ \nonumber 
  & & \qquad\qquad\qquad+(34\rho^2+36a^2)\log\rho +(10\rho^2+63a^2)\log(\rho^2+9a^2)\Big]
\end{eqnarray}
The five--form flux is as usual given by 
\begin{equation}
  \hat{F}_5 \,=\, (1+\tilde{*}_{10})(d\hat h^{-1}\wedge d^4x)\,.
\end{equation}
In order to solve the supergravity equations of motion, the warp factor has to fulfill
\begin{equation}
  \hat h^2\,\Delta \hat h^{-1}-2 \hat h^3\,\partial_m \hat h^{-1}\,\partial_n \hat h^{-1} g^{mn}\,=\,-\Delta\hat{h}
    \,=\,*_6 \left(\frac{\hat G_3\wedge \overline{\hat G}_3}{6\left(\overline{\tau}-\tau\right)}\right)\,=\,\frac{1}{6}*_6 d\hat{F}_5\,,
\end{equation}
where $\Delta$ is the Laplacian on the unwarped resolved conifold and all indices are raised and lowered with the unwarped metric.
This should be evaluated in linear order in N, since we solved the SuGra eom for the fluxes also in linear order. However, we were unable to find an analytic solution to this problem, so we need to employ some simplification. We can take the limit $\rho\gg a$, i.e. we restrict ourselves to be far from the tip. As in the limit $a\to 0$ we recover the singular conifold setup \cite{ouyang}, we know our solution takes the form 
\begin{eqnarray}
  \hat h(\rho,\theta_1,\theta_2) &=& 1+\frac{L^4}{r^4}\,\left\{1+\frac{24g_sP^2}{\pi\alpha'Q}\,\log\rho\left[1+\frac{3g_sN}{2\pi\alpha'}
    \left(\log\rho+\frac{1}{2}\right)\right.\right.\\ \nonumber
  && \left.\left. \qquad\qquad\qquad\qquad +\frac{g_sN}{2\pi\alpha'}\,\log\left(\sin\frac{\theta_1}{2}\sin\frac{\theta_2}{2}\right)\right]\right\} 
    +\mathcal{O}\left(\frac{a^2}{\rho^2}\right)
\end{eqnarray}
with $L^4=27\pi g_s\alpha'Q/4$. Unfortunately, we cannot give an explicit expression for the $a^2/\rho^2$ corrections. However, above result is already an improvement over using the simple Klebanov--Tseytlin warp factor (which is strictly only valid for the singular solution, but is often employed in the deformed KS geometry).

\end{appendix}


\bibliographystyle{utphys}

\bibliography{dbraneinfl}

\end{document}